\documentclass[a4paper, 11pt]{article}
\pdfoutput=1

\usepackage{jheppub}
\usepackage{afterpage}
\usepackage{comment}

\usepackage{float}
\usepackage{graphicx}
\usepackage{caption}
\usepackage{subcaption}

\usepackage{tikz,xcolor,hyperref}

\definecolor{lime}{HTML}{A6CE39}
\DeclareRobustCommand{\orcidicon}{\hspace{-1.5mm}
	\begin{tikzpicture}
		\draw[lime, fill=lime] (0,0) 
		circle [radius=0.12] 
		node[white] {{\fontfamily{qag}\selectfont \tiny \,ID}};
		\draw[white, fill=white] (-0.0525,0.095) 
		circle [radius=0.007];
	\end{tikzpicture}
	\hspace{-3.5mm}
}

\foreach \x in {A, ..., Z}{\expandafter\xdef\csname orcid\x\endcsname{\noexpand\href{https://orcid.org/\csname orcidauthor\x\endcsname}
		{\noexpand\orcidicon}}
}

\allowdisplaybreaks

\title{Analytic treatment of 3-flavor neutrino oscillation and decay in matter}

\author[a]{Dibya~S.~Chattopadhyay\orcidA{},}
\affiliation[a]{Tata Institute of Fundamental Research, Homi Bhabha Road, Colaba, Mumbai 400005, India}
\emailAdd{d.s.chattopadhyay@theory.tifr.res.in}

\author[b,c,d]{ Kaustav~Chakraborty\orcidB{},}
\affiliation[b]{Physical Research Laboratory, Navrangpura, Ahmedabad 380009, India}
\affiliation[c]{Tsung-Dao Lee Institute \& School of Physics and Astronomy, Shanghai Jiao Tong University, Shanghai 200240, China}
\affiliation[d]{Key Laboratory for Particle Astrophysics and Cosmology (MOE) \& Shanghai Key Laboratory for Particle Physics and Cosmology, Shanghai Jiao Tong University, Shanghai 200240, China}
\emailAdd{kaustav.chk@gmail.com}

\author[a]{ Amol~Dighe\orcidC{},}
\emailAdd{amol@theory.tifr.res.in}

\author[b]{ Srubabati~Goswami\orcidD{}}
\emailAdd{sruba@prl.res.in}

\abstract{We present compact analytic expressions for 3-flavor neutrino oscillation probabilities with invisible neutrino decay, where matter effects have been explicitly included.
We take into account the possibility that the oscillation and decay components of the effective Hamiltonian do not commute.
This is achieved by employing the techniques of inverse Baker-Campbell-Hausdorff (BCH) expansion and the Cayley-Hamilton theorem applied in the 3-flavor framework.
If only the vacuum mass eigenstate $\nu_3$ decays, we show that the treatment of neutrino propagation may be reduced to an effective 2-flavor analysis in the One Mass Scale Dominance (OMSD) approximation.
The oscillation probabilities for $P_{\mu\mu}$, $P_{ee}$, $P_{e\mu}$ and $P_{\mu e}$ --- relevant for reactor, long baseline and atmospheric neutrino experiments --- are obtained as perturbative expansions for the case of only $\nu_3$ decay, as well as for the more general scenario where all components of the decay matrix are non-zero.
The analytic results thus obtained match the exact numerical results for constant density matter to a high precision and provide physical insights into possible effects of the decay of neutrinos as they propagate through Earth matter.
We find that the effects of neutrino decay are most likely to be observable in $P_{\mu\mu}$.
We also point out that at any long baseline, the oscillation dips in $P_{\mu\mu}$ can show higher survival probabilities in the case with decay than without decay, and explain this feature using our analytic approximations.
	}

\preprint{TIFR/TH/22-17}


\begin{document}

\maketitle

\section{Introduction}
\label{sec:intro}
The observations of neutrino oscillations in terrestrial experiments have 
conclusively established that neutrinos have tiny but non-zero masses~\cite{ParticleDataGroup:2020ssz}.  The past and present neutrino oscillation experiments have consolidated the 
3-flavor paradigm of neutrino mixing, measuring most of the parameters governing the oscillations with a considerable precision~\cite{deSalas:2020pgw,Esteban:2020cvm,Capozzi:2021fjo}. The parameters that are quite well measured include the solar mass-squared difference ($\Delta m^2_{21}$), the magnitude of the atmospheric mass-squared difference ($|\Delta m^2_{31}|$) and the mixing angles ($\theta_{12}$, $\theta_{23}$, $\theta_{13}$).
Future  experiments are aimed to unambiguously determine the  unknown oscillation parameters --- the sign of $\Delta m^2_{31}$, the octant of the atmospheric neutrino mixing angle $\theta_{23}$, and the CP phase $\delta_{\text{CP}}$ in the lepton sector.  In addition, these experiments can also probe physics beyond the Standard Model (BSM), 
which can give rise to sub-leading effects on top of the dominant neutrino oscillation  phenomenon.  

A possible BSM scenario is the invisible  decay of neutrinos to final states which do not interact with the detector.  
Such decays can occur in two ways: For Dirac neutrinos,  the decay mode is  $\nu_{j}\rightarrow\bar{\nu}_{iR}+\chi$, 
where $\bar{\nu}_{iR}$ is a singlet fermion  and $\chi$ is an iso-singlet scalar 
\cite{Acker:1991ej,Acker:1993sz}.  For Majorana neutrinos, the decay channel is $\nu_{j}\rightarrow\nu_{s}+J$, where $\nu_s$ is a sterile neutrino and  $J$  is  a Majoron 
\cite{Gelmini:1980re, Chikashige:1980ui}.  The LEP constraints on the decay of the Z boson  to invisible particles disfavor the triplet Majoron model, indicating that this Majoron would be primarily a singlet \cite{Pakvasa:1999ta}.
The solar and atmospheric neutrino data disfavor pure neutrino decay solutions \cite{Bahcall:1972my,Bahcall:1986gq,LoSecco:1998cd} to these anomalies.  
However, combined oscillation and decay scenarios, with decay as a sub-leading effect, can still be allowed and have been studied in the context of solar  
\cite{Berezhiani:1987gf,Berezhiani:1991vk,Berezhiani:1992xg,Choubey:2000an,Bandyopadhyay:2001ct,Joshipura:2002fb,Bandyopadhyay:2002qg,Picoreti:2015ika} and atmospheric  neutrinos \cite{Barger:1998xk,Fogli:1999qt, Choubey:1999ir, Barger:1999bg}.   
The decay effects for the neutrino mass eigenstate $\nu_i$ with mass $m_i$ and lifetime $\tau_i$ are typically characterized by the term  $\exp{[ - (m_i/\tau_i) (L/ E_\nu)]}$, which represents the fraction of neutrinos 
of energy $E_\nu$ that decay after traversing a distance $L$.

Constraints on invisible neutrino decay have been obtained from several neutrino observations.
Assuming only the state $\nu_2$ decays, a lower bound on the decay lifetime has been obtained from the solar neutrino data to be
$\tau_{2}/m_{2} > 8.7\times10^{-5}$ s/eV (99\% C.L.)~\cite{Bandyopadhyay:2002qg}.  Considering  the  possibility of  both $\nu_2$ and $\nu_1$ undergoing 
invisible decay, and combining high and low energy solar neutrino data along with KamLAND reactor data, the limits are
$\tau_1/m_1 > 4 \times 10^{-3}$ s/eV and $\tau_2/m_2 > 7 \times 10^{-4}$ 
s/eV  at  95\%  C.L.~\cite{Berryman:2014qha}. 
The constraints obtained from the observation of supernova SN1987A neutrinos~\cite{Frieman:1987as} imply that for at least one neutrino mass eigenstate, $\tau/m > 10^{5}$ s/eV.
The global analysis of atmospheric and long-baseline neutrino experiments in the context of decay plus oscillation 
solutions  to neutrino anomalies \cite{GonzalezGarcia:2008ru} puts the bound $\tau_3 /m_3 > 2.9 \times 10^{-10}$ s/eV (90\% C.L.).
Analysis of oscillation plus decay scenario  for  MINOS and T2K \cite{Gomes:2014yua}  and  more recently in the context of T2K and NO$\nu$A data \cite{Choubey:2018cfz} constrain $\tau_3 /m_3 > 2.8 \times 10^{-12}$  s/eV (90\% C.L.) and $\tau_3 /m_3 > 1.5 \times 10^{-12}$ s/eV ($3\sigma$), respectively.

Prospects of constraining neutrino decay in future high precision  accelerator experiments  like DUNE, T2HK/T2HKK, ESS$\nu$SB have been examined in~\cite{Chakraborty:2020cfu}. Effects of decay plus oscillation in the context of the JUNO experiment have been studied 
in~\cite{Abrahao:2015rba}.  Invisible decay combined with oscillation in the context of future atmospheric neutrino experiments have been studied  
in~\cite{Choubey:2017eyg} for INO, and in~\cite{deSalas:2018kri} for  KM3Net-ORCA. 
Invisible neutrino decay has also been studied in the context of neutrino telescopes \cite{Beacom:2002cb, Pakvasa:2012db, Denton:2018aml}. 
Strong bounds on invisible neutrino decay lifetime can come from cosmology \cite{Hannestad:2005ex, Basboll:2008fx, Escudero:2019gfk}.  Recent studies on constraining  invisible neutrino 
decay from precision cosmology has been performed in~\cite{ Barenboim:2020vrr,Chen:2022idm}. 

Many of the scenarios with oscillation and decay studied in the  above references were explored prior to  the discovery of non-zero $\theta_{13}$, and 
considered neutrino decay in a 2-flavor scenario in vacuum. Notable exceptions like~\cite{Lindner:2001fx} and some of the more recent works~\cite{Abrahao:2015rba,Ghoshal_2021} have calculated the expressions for combined oscillation and decay probabilities in vacuum in the 3-flavor scenario. However, these studies assume that the neutrino mass eigenstates in vacuum are the same as the decay eigenstates. Compact analytic expressions for 2-flavor mixing in vacuum, which do not make this assumption and use a non-unitary matrix for the diagonalization of the Hamiltonian, have been presented in~\cite{BERRYMAN201574}.
In all the above studies, matter effects, if relevant, have been included in numerical analyses.
No study offering compact analytic expressions for probabilities in matter in the presence of invisible decay had been accomplished till recently, even in the simple case of 2-flavor mixing.
In a recent paper~\cite{Chattopadhyay:2021eba}, we have presented the first-ever analytic expressions for probabilities with 2-flavor neutrino oscillation and decay in matter. 

The propagation of decaying neutrinos can be described in terms of a non-Hermitian Hamiltonian, where the Hermitian component drives the  neutrino oscillations, and the anti-Hermitian component governs the neutrino decay. This may be written as
\begin{equation}
	\mathcal{H}=H-i \Gamma/2\;,
	\label{eq:Heff}
\end{equation}
where $H$ is the Hermitian component and $-i\Gamma/2$ is the anti-Hermitian component of the total Hamiltonian $\mathcal{H}$.
In terms of ${\mathcal H}$, the flavor evolution
of neutrinos takes the form
\begin{equation}{\label{eq:Generalevolution}}
	[\nu(t)]_\alpha = [e^{-i \mathcal{H} t}]_{\alpha \beta}\; [\nu(0)]_\beta \; ,
\end{equation}
where, $[...]$ denotes a matrix.
While eq.~(\ref{eq:Heff}) is valid in all bases, a convenient basis is that where the Hermitian part of the Hamiltonian is diagonal \cite{Chattopadhyay:2021eba}. This corresponds to the basis of mass
eigenstates in matter, in the absence of decay.
The matrix $H$ expressed in this basis ($H_m$) is a diagonal matrix whose elements depend on
neutrino mass squared differences, neutrino energy, and the Earth matter potential. 
On the other hand,  the matrix $\Gamma$ in this basis ($\Gamma_m$) may be non-diagonal.
Thus, in  general, $H_m$ and $\Gamma_m$ do not commute, and hence cannot be
simultaneously diagonalized.
Since $[H_m, \Gamma_m] \neq 0$ in general, $\mathcal{H}_m$ is not a normal
matrix, and 
\begin{equation}
	e^{- i {\mathcal H}_m t} \neq e^{- i H_m t}  e^{- \Gamma_m t/2} \; .
	\label{eq:naiveBCH}
\end{equation}
Thus, in order to calculate the oscillation amplitude matrix, one has to express the left hand side of eq.~(\ref{eq:naiveBCH}) in terms of a chain of
commutators using the inverse Baker-Campbell-Hausdorff (BCH) formula,
also known as the Zassenhaus formula \cite{https://doi.org/10.1002/cpa.3160070404,CASAS20122386}.
In~\cite{Chattopadhyay:2021eba}, we have developed a procedure for calculating the neutrino probabilities for combined oscillation and decay  in constant density matter for two neutrino flavors,
employing a resummation procedure for the Zassenhaus expansion~\cite{Kimura:2017xxz}.

In this paper, we set out to derive the analytic probabilities in constant density matter  for the 3-flavor mixing. It may be recalled here that even in the absence of decay it is difficult to deal with three-neutrino propagation in matter, due to the mismatch between the propagation and the interaction eigenstates of neutrinos. 
In fact, no exact compact analytic solution exists. However, one can derive the probabilities perturbatively, as series expansions in terms 
of small parameters that have been identified.  Two approximations that are often used are (i) the One Mass Scale Dominance (OMSD) approximation which exploits the  
fact that the solar mass-squared difference  is much smaller than the  atmospheric  one, i.e. $\Delta m^2_{21} << \Delta m^2_{31}$,
and takes $\Delta m^2_{21}= 0$,
(ii) the $\alpha-s_{13}$ expansion in terms of small parameters $\alpha \equiv  \Delta m^2_{21}/ \Delta m^2_{31}$
and $s_{13} \equiv \sin (\theta_{13})$ \cite{Akhmedov:2004ny}.
The inclusion of decay into the picture makes this problem even more complicated.

Additional complications with decaying neutrinos arise because a non-Hermitian effective Hamiltonian cannot be diagonalized using  a unitary matrix~\cite{BERRYMAN201574}.
	The mismatch between the decay and 
	mass eigenstates in matter introduces additional  parameters.  For instance, even if one starts with a single decaying vacuum mass eigenstate,
	all the mass eigenstates in matter will develop decaying components.
	Such non-intuitive features are brought out in our analytic treatment.
	We compute  the  neutrino conversion/survival probabilities, in vacuum as well as in the presence of matter, using different techniques and under various approximations.
	
	In section~\ref{sec:hamiltonian}, we present our parametrization of the effective Hamiltonian corresponding to neutrino oscillation and decay, and formally set up the scheme for perturbative expansion in small parameters. We consider two scenarios, one where only the mass eigenstate $\nu_3$ in vacuum decays, and the more general one, where all components of the decay matrix $\Gamma$ in the vacuum mass basis are non-zero. The former is motivated by the observation that the limits on the decay lifetimes of $\nu_1$ and $\nu_2$ are quite stringent. The latter scenario is also of relevance, since exotic interactions of neutrinos in matter may relax some of the constraints previously discussed. The rest of the paper is organized as follows.
	
	\begin{itemize}
		\item[(i)] We first calculate in section~\ref{sec:OMSD} the results in the OMSD approximation, in the scenario where only the third mass eigenstates  $\nu_3$ decays  in vacuum.
		We  show that the 2-flavor  formalism developed in~\cite{Chattopadhyay:2021eba} can be used  effectively in this scenario.
		
		\item[(ii)] We then develop in section~\ref{sec:zassenhaus} the 3-flavor Zassenhaus expansion for $\nu_3$ decaying in vacuum, relaxing the assumption $\Delta m^2_{21}=0$. 
		In this case, the probabilities in vacuum are explicitly presented in terms of  a series expansion  in   $\alpha$,  the 
		reactor mixing angle $\sin  \theta_{13}$ as well as the decay parameter $\gamma_3$ . 
		We also give the prescription for obtaining the probabilities in matter using the Zassenhaus expansion in terms of  $\alpha_m$, $s_{13}^m$, 
		$\gamma_{i}^m$ and the mismatch parameters $\gamma_{ij}^m$, where the suffix `$m$' denotes the corresponding quantities in matter.  
		
		\item[(iii)] We bring out the explicit matter dependence in section~\ref{sec:CayleyHamiltonnu3} and section~\ref{sec:CayleyHamiltonGamma} by using the 
		Cayley-Hamilton theorem, which allows us to write down the probabilities in terms of parameters defined in vacuum, with the dependence on matter potential expressed explicitly.
		The scenario with only $\nu_3$ decay is considered in section~\ref{sec:CayleyHamiltonnu3}, whereas the general decay matrix $\Gamma$ (with all its components non-zero) is considered in section~\ref{sec:CayleyHamiltonGamma}.
	\end{itemize}
In section~\ref{sec:numcom}, we compare the analytic expansions against numerically computed probabilities (in the PREM-averaged constant matter density along the line of propagation~\cite{Dziewonski:1981xy,Gandhi:2004bj}) for baselines of 1300 km and 7000 km. We further extend our analysis to a wider range of baselines and identify the regions in the $(E_\nu,\, L)$ parameter space where the accuracy of our analytically calculated probabilities is very high. Our study reveals some non-intuitive features of the neutrino oscillation probabilities, which can be explained with the analytic expressions. We end in section~\ref{sec:summary} with summary and conclusions. 

\section{The effective Hamiltonian}
\label{sec:hamiltonian}
Neutrino decay may be analyzed in terms of an effective Hamiltonian which is non-Hermitian. Such a Hamiltonian can be expressed as a combination of Hermitian and anti-Hermitian components, which describe the neutrino oscillation and decay, respectively. 
In the Weisskopf-Wigner approximation~\cite{kabir1968cp,PhysRevD.1.2683},
one may write the decay matrix as
\begin{align}
	\Gamma_{ij} =&\; 2\pi \sum_{k} \langle \nu_{i} |\mathcal{H}^\prime| \phi_k
	\rangle \langle \phi_k |\mathcal{H}^\prime| \nu_j \rangle \,\delta(E_k-E_\nu)
	\; .
	\label{eq:Gammaij}
\end{align}
Here $|\phi_k \rangle$ represents the final states with energy $E_k$ to which neutrinos decay, and $\mathcal{H}^\prime$ is the interaction term due to BSM physics.

In principle, all the elements of the matrices $H_{ij}$ and $\Gamma_{ij}$ can be non-zero in any generic basis.
For the simplicity of discussion, we can write the effective Hamiltonian in the basis where the Hermitian component of the Hamiltonian is diagonal, and use the suffix `$m$' for the corresponding matrix. Note that the suffix `$m$' emphasizes that this form is applicable even in the presence of matter. In this basis, the Hamiltonian takes the form
\begin{equation}
	\label{eq:gendecay}
	\mathcal{H}_m \equiv H_m-\frac{i}{2} \Gamma_m \equiv \left(
	\begin{array}{ccc}
		a_1& 0 & 0\\
		0& a_2 & 0 \\
		0 & 0& a_3 \\
	\end{array}
	\right)- \frac{i}{2} \left(
	\begin{array}{ccc}
		2 b_1 & b_{12} e^{i \text{$\chi_{12} $}} &b_{13}  e^{i \text{$\chi_{13} $}} \\
		b_{12}  e^{-i \text{$\chi_{12} $}} & 2 b_2 & b_{23} e^{i \text{$\chi_{23} $}} \\
		 b_{13}  e^{-i \text{$\chi_{13} $}} & b_{23}  e^{-i \text{$\chi_{23} $}} &2 b_3 \\
	\end{array}
	\right).
\end{equation}
Here $H_m$ and $\Gamma_m$ are Hermitian matrices, with $a_i$'s being the eigenvalues of the Hermitian component $H_m$ that are responsible for neutrino oscillations. The $b_i$'s and $b_{ij}$'s are the elements of the decay matrix, which may be obtained from eq.~(\ref{eq:Gammaij}) by using the corresponding basis.
This is a 3-flavor generalization of the convention introduced in~\cite{Chattopadhyay:2021eba}. Note that all the $a_i$'s, $b_i$'s, $b_{ij}$'s and $\chi_{ij}$'s are real. Since the effects of decay is expected to be sub-dominant compared to the oscillation effects, $O(b_i),\, O(b_{ij})<O(a_i)$. The values of $b_i$'s and  $b_{ij}$'s are further constrained by the condition that the decay matrix, $\Gamma_m$ is positive definite. 

Most of the earlier literature analyzing neutrino decay only considers scenario where decay eigenstates are the same as mass eigenstates, i.e. the eigenvectors of $H_m$ and $\Gamma_m$ coincide, which corresponds to $\gamma_{ij}=0$.  However, this scenario is clearly not applicable if $[H_m,\Gamma_m] \neq 0$. Further, due to earth matter effects, the eigenstates of $H_m$ are in general different from the neutrino mass eigenstates in vacuum. Therefore, even in the scenario where mass eigenstate and decay eigenstates are identical in vacuum, the off-diagonal decay terms $\gamma_{ij}$ will invariably arise in matter. This makes it imperative to incorporate the effects of $\gamma_{ij}$'s, while calculating the analytic expressions for neutrino probabilities.

In section~\ref{sec:hamiltonian_nu3}, we first separately discuss the  Hamiltonian for the scenario where only one mass eigenstate in vacuum, $\nu_3$, decays. This corresponds to the scenario primarily considered in literature. Note that this does not mean that the decay matrix $\Gamma_m$ in the effective mass basis in matter, as defined in eq.~(\ref{eq:gendecay}), will only have one non-zero component. In section~\ref{sec:hamiltonian_gen}, we generalize to the scenario where all elements of $\Gamma$ can be non-zero even in vacuum.

\subsection{Decay of $\nu_3$ only}
\label{sec:hamiltonian_nu3}
In the scenario where only the $\nu_3$ mass eigenstate in vacuum decays, the neutrino propagation in the presence of matter may be described in terms of the effective Hamiltonian
\begin{align}
	\mathcal{H}_{f}^{(\gamma_3)}
	=&\; \frac{1}{2 E_{\nu }} U \left[ \left(
	\begin{array}{ccc}
		0 & 0 & 0 \\
		0 & \text{$\Delta $m}_{21}^2 & 0 \\
		0 & 0 & \text{$\Delta $m}_{31}^2 \\
	\end{array}
	\right)-i \left(
	\begin{array}{ccc}
		0\quad & 0 \;& 0\; \\
		0\quad & 0 \;& 0 \;\\
		0 \quad& 0 \;& \gamma _3 \text{$\Delta $m}_{31}^2  \\
	\end{array}
	\right) \right] U^\dagger
	 +\left(
	\begin{array}{ccc}
		V_{\text{cc}} & 0\; & 0 \\
		0 & 0 &\; 0 \\
		0 & 0 &\; 0 \\
	\end{array}
	\right),
	\label{eq:Hnu3decay}
\end{align}
where $\Delta m_{21}^2 \equiv m_2^2 - m_1^2$, $\Delta m_{31}^2 \equiv m_3^2 - m_1^2$, and $V_{\text{cc}}=\sqrt{2} G_F N_e$, with $G_F$ the Fermi constant, and $N_e$ the electron number density. Here $\gamma_3$ is defined such that $ \gamma_3 \Delta m^2_{31} = m_3/\tau_3$, where $m_3$ is the mass, and $\tau_3$ is the lifetime of $\nu_3$. The neutrino mixing matrix $U=U_{23}\, U_{13}\, U_{12}$, with
\begin{equation}
	U_{23}=\left(
	\begin{array}{ccc}
		1 & 0 & 0 \\
		0 & c_{23} & s_{23} \\
		0 & -s_{23}& c_{23}\\
	\end{array}
	\right),\;
	U_{13}=\left(
	\begin{array}{ccc}
		c_{13} & 0 & s_{13} e^{-i \delta _{\text{CP}}} \\
		0 & 1 & 0 \\
		- s_{13} e^{i \delta _{\text{CP}}} & 0 & c_{13} \\
	\end{array}
	\right),\; U_{12}=\left(
	\begin{array}{ccc}
		c_{12} & s_{12}& \; 0 \\
		-s_{12} & c_{12} & \; 0 \\
		0 & 0 & \; 1 \\
	\end{array}
	\right).
\end{equation}
We use the notation $s_{ij}\equiv \sin\theta_{ij}$ and $c_{ij}\equiv \cos\theta_{ij}$ throughout the rest of this paper.

Since the exact compact expressions for neutrino oscillation probabilities in the 3-flavor are impossible to find, we will be using perturbation techniques in this paper. The small dimensionless parameters in which the perturbative expansion can be carried out are $\alpha$, $s_{13}$ and $\gamma_3$. For convenience, we express all of these in terms of a common book-keeping small parameter $\lambda \equiv 0.2$. Based on their magnitudes or current upper limits on the values of these parameters, we assign them the powers of $\lambda$ as
\begin{align}
		\alpha \approx 0.03 \simeq O(\lambda ^2)\;,\qquad
		s_{13}\simeq  0.14 \simeq O(\lambda)\;,\qquad
		\gamma _3 \lesssim 0.1 \simeq O(\lambda)\;.
\end{align}
We also define the dimensionless quantities
\begin{equation}
	A=\frac{2E_\nu V_{\text{cc}}}{\Delta m_{31}^2}\;,\qquad \qquad \Delta =\frac{\Delta m_{31}^2 L}{4 E_\nu}\;,
\end{equation}
since these are combinations that often appear in the final expressions.

Note that the addition of the matter effect term $V_{cc}$ mixes the eigenstate $\nu_3$ with the others. This implies that when neutrinos propagate through matter, the decay matrix $\Gamma_m$ in the effective mass basis in matter, as defined in eq.~(\ref{eq:gendecay}), would have non-zero elements other than $\gamma_3^m$.
As a result, more than one neutrino mass eigenstates in matter will now be seen to undergo decay.

\subsection{General decay matrix $\Gamma$}
\label{sec:hamiltonian_gen}
The effective Hamiltonian in flavor basis, in the presence of matter, may be expressed in the form
\begin{align}
	\label{eq:Hgendecay}
	\mathcal{H}_{f}^{(\Gamma)}=&\; U \left[ \frac{1}{2 E_{\nu }} \left(
	\begin{array}{ccc}
		0 & 0 & 0 \\
		0 & \text{$\Delta $m}_{21}^2 & 0 \\
		0 & 0 & \text{$\Delta $m}_{31}^2 \\
	\end{array}
	\right)-\frac{i}{2} \Gamma \right] U^\dagger
	+\left(
	\begin{array}{ccc}
		V_{\text{cc}} & 0\; & 0 \\
		0 & 0 &\; 0 \\
		0 & 0 &\; 0 \\
	\end{array}
	\right)\;,
\end{align}
where
\begin{equation}
	\Gamma=\frac{\Delta m_{31}^2}{E_\nu}\left(
	\begin{array}{ccc}
		\gamma_1 & \frac{1}{2}\gamma_{12}  e^{i \chi _{12}} &  \frac{1}{2}\gamma_{13} e^{i \chi _{13}}  \\
		\frac{1}{2}\gamma_{12}  e^{-i \chi _{12}}  & \gamma_2 &  \frac{1}{2}\gamma_{23}  e^{i \chi _{23}} \\
		\frac{1}{2}\gamma_{13} e^{-i \chi _{13}} & \frac{1}{2}\gamma_{23}  e^{-i \chi _{23}} & \gamma_3 \\
	\end{array}
	\right)\;.
\end{equation}
Since neutrino decay has not yet been observed in reactor or long-baseline experiments, the length-scales associated with decay can safely be assumed to be larger by at least a factor of $\sim 1/O(\lambda)$ than the corresponding oscillation length-scales. This would imply
\begin{equation}
	\gamma_1 \Delta m^2_{31}\lesssim \; O(\lambda) \Delta m^2_{21}\;,\qquad \; \gamma_2 \Delta m^2_{31}\; \lesssim \; O(\lambda) \Delta m^2_{21}\;,\qquad \gamma_3 \Delta m^2_{31} \lesssim O(\lambda) \Delta m^2_{31}\;,
\end{equation}
and therefore,
\begin{equation}
	\gamma_{1} \lesssim O(\lambda^3)\;,\qquad  \gamma_{2} \lesssim O(\lambda^3)\;,\qquad \gamma_3\lesssim O(\lambda)\;.
\end{equation}
Furthermore, since the decay matrix $\Gamma$ must be positive definite, its off-diagonal terms are constrained to be $O(\gamma_{ij}^2) \lesssim O(\gamma_i ) O(\gamma_j)$. 
Each of the dimensionless quantities for the general decay scenario are now expressed in terms of powers of $\lambda$. For our calculations, we take
\begin{equation}
	\gamma_{1}\,,\, \gamma_2 \sim O(\lambda^3)\;,\qquad \gamma_3 \sim O(\lambda)\;,\qquad \gamma_{12} \sim O(\lambda^3)\;,\qquad \gamma_{13}\, ,\,\gamma_{23} \sim O(\lambda^2)\;,
\end{equation}
which allows the values of $\gamma$ as large as permitted by constraints above. In summary, we take
\begin{equation}
	\Gamma \sim \frac{\Delta m_{31}^2}{E_\nu}\left(
	\begin{array}{ccc}
		\lambda^3 & \lambda^3& \lambda^2 \\
		\lambda^3& \lambda^3 & \lambda^2 \\
		\lambda^2&\lambda^2& \lambda \\
	\end{array}
	\right).
\end{equation}

For future long baseline neutrino experiments with the expected absolute accuracy in the probabilities $\sim 1\,\%$, it is enough to have analytic expressions accurate up to $O(\lambda^3)$. Therefore, in our analytic calculations we ignore any $\gamma_{ij}^2$ terms, as well as $\gamma_1^2$ and $\gamma_2^2$ terms.
Naively, one may expect the effects of $\nu_3$ decay to be dominant in all cases. However, we shall see later explicitly that there are cases where the other $\gamma_i,\; \gamma_{ij}$ elements can contribute to the same extent.

In the next sections, we'll calculate analytic expressions for neutrino conversion/ survival probabilities, in the presence of matter and decay, as perturbative expansions in powers of the book-keeping parameter $\lambda$. 
\section{One Mass Scale Dominance (OMSD), with decay of $\nu_3$ only}
\label{sec:OMSD}
In the OMSD approximation, using the observation that $\alpha \sim O(\lambda^2)$, the effect of $\Delta m^2_{21}$ is neglected with respect to $\Delta m^2_{31}$.
The propagation can then be studied in an effective 2-flavor approximation in the $1-3$ sector, when only the $\nu_3$ mass eigenstate in vacuum decays.
As a result, the Hamiltonian becomes independent of the mixing angle $\theta_{12}$ and the CP phase $\delta_{CP}$.
\begin{equation}
\mathcal{H}_{f}^{(\text{OMSD})}=\frac{\text{$\Delta $m}_{31}^2}{2 E_{\nu }}\left[ U \left(
\begin{array}{ccc}
0\quad & 0 & 0 \\
0\quad & 0 & 0 \\
0\quad & 0 & \;1-i\gamma_3 \\
\end{array}
\right) U^\dagger+\left(
\begin{array}{ccc}
A \; & 0 \; & 0 \; \\
0 \; & 0 \; & 0 \; \\
0 \; & 0 \; & 0 \; \\
\end{array}
\right)\right]\;,
\label{omsd-hamiltonian}
\end{equation}
where the mixing matrix $U$ has the simplified form
\begin{eqnarray}
U = R_{23}(\theta_{23}) \cdot R_{13}(\theta_{13}) = 
\begin{pmatrix}
1 & 0 & 0 \\
0 & c_{23} & s_{23} \\
0 & -s_{23} & c_{23} \\
\end{pmatrix}
\begin{pmatrix}
c_{13} & 0 & s_{13} \\
0 & 1 & 0 \\
-s_{13} & 0 & c_{13} \\
\end{pmatrix}. 
\label{omsd-mix-mat}
\end{eqnarray}
The matter potential matrix in eq.~(\ref{omsd-hamiltonian}) commutes with the rotation matrix $R_{23}$ in eq.~(\ref{omsd-mix-mat}). Therefore, the matter effects only modify the $R_{13}$ part of the mixing matrix. 
We go to a basis rotated by $R_{23}$ in which the  effective Hamiltonian takes the form
\begin{eqnarray}
\label{eq:tildehamiltonian}
\widetilde{\mathcal{H}}_{f}^{(\text{OMSD})}= \dfrac{\Delta m^2_{31}}{2E_\nu}\left[ R_{13}
\begin{pmatrix}
0 \qquad & 0 \quad & 0 \\
0 \qquad & 0 \quad & 0 \\
0 \qquad & 0 \quad & 1-i\gamma_3 \\
\end{pmatrix} 
R_{13}^\dagger
+
\begin{pmatrix}
A \quad& 0 \quad& 0 \\
0 \quad& 0 \quad & 0 \\
0 \quad& 0 \quad & 0 \\
\end{pmatrix} \right].
\end{eqnarray}
We represent by ``$\sim$" the quantities in this $R_{23}$-rotated basis.
The Hamiltonian in eq.~(\ref{eq:tildehamiltonian}) can be decomposed into a Hermitian matrix responsible for oscillatory behavior, and an anti-Hermitian matrix 
responsible for decay; represented as
\begin{equation}
	\widetilde{\mathcal{H}}_{f}^{(\text{OMSD})}=\widetilde{H}_{f}^{(\text{OMSD})} -\frac{i}{2}\widetilde{\Gamma}^{\text{(OMSD)}}_f.
\end{equation}
Here,
\begin{equation}
	\widetilde{\Gamma}^{\text{(OMSD)}}_f\equiv \frac{\Delta m^2_{31}}{E_\nu} R_{13}\, \text{diag}[(0,0,\gamma_3) ]R_{13}^\dagger\;.
\end{equation}
Now, the Hermitian matrix $\widetilde{H}_{f}^{(\text{OMSD})}$ can be diagonalized as
\begin{equation}
\widetilde{H}_{f}^{(\text{OMSD})}=\dfrac{\Delta m^2_{31}}{2E_\nu} R_{13}^m \begin{pmatrix}
\Lambda_1 & 0 & 0 \\
0 & \Lambda_2 & 0 \\
0 & 0 & \Lambda_3 \\
\end{pmatrix} R^{m \; \dagger}_{13}\;.
\end{equation}
The eigenvalues $\Lambda_i$ can be obtained as
\begin{equation}
\Lambda_{1,3} = \frac{\Delta m^2_{31}}{4E}\left[1 + A \mp C_{13} \right]\;, \qquad\Lambda_2 = 0\;,
\end{equation}
where $C_{13}\equiv \sqrt{( \cos2\theta_{13} - A)^2 + ( \sin2\theta_{13})^2}\,$. The  rotation matrix $R_{13}^m = R_{13}(\theta_{13}^m)$, where
\begin{eqnarray}
	\tan 2\theta_{13}^m = \frac{ \sin 2\theta_{13}}{ \cos 2\theta_{13} - A}\;.
\end{eqnarray}
A rotation via $R_{13}^m$ takes us to the ``OMSD basis'' where the total Hamiltonian can be represented as
\begin{equation}
\widetilde{\mathcal{H}}^{(\text{OMSD})}_{m}=  \text{diag}[(\Lambda_1,\Lambda_2,\Lambda_3)]-\frac{i}{2} \widetilde{\Gamma}^{\text{(OMSD)}}_{m}\;.
\end{equation} 
Here, 
\begin{eqnarray}
\widetilde{\Gamma}^{\text{(OMSD)}}_{m}=R_{13}^{m\; \dagger}\,\widetilde{\Gamma}^{\text{(OMSD)}}_f\, R_{13}^m = \frac{\Delta m^2_{31}}{ E_\nu} 
	\begin{pmatrix}
		\gamma_{1}^m & 0 & \frac{1}{2}\gamma_{13}^m \\
		0 & 0 & 0 \\
		\frac{1}{2} \gamma_{13}^m & 0 & \gamma_{3}^m \\
	\end{pmatrix},
\end{eqnarray}
with
\begin{equation}
	\gamma_{1}^m \equiv \gamma _3 \sin ^2 \delta \theta \;,\qquad \gamma_{3}^m \equiv \gamma _3 \cos ^2 \delta \theta \;,\qquad \gamma_{13}^m \equiv  -\gamma_3 \sin (2\, \delta\theta)\;,
\end{equation} 
where $\delta \theta \equiv \theta_{13}^m-\theta_{13}$. 
Note that, even though we started with only $\nu_3$ decaying, the decay matrix in the OMSD basis has multiple non-zero diagonal as well as off-diagonal elements.
Since the second row and column of this matrix is zero and since $\Lambda_2$ is vanishing, the 2-flavor technique developed in~\cite{Chattopadhyay:2021eba} can be directly applied.
In the next subsection, we calculate the neutrino survival/ conversion probabilities $P_{\mu\mu}$, $P_{ee}$, $P_{e\mu}$ and $P_{\mu e}$ using the OMSD approximation.

\subsection{Neutrino probabilities with OMSD approximation}
The amplitude for $\nu_\alpha \to \nu_\beta$ in the OMSD approximation is given by
\begin{equation}
	A(\nu_\alpha\to\nu_\beta) =[e^{-i \mathcal{H}_{f}^{(\text{OMSD})} L}]_{\beta \alpha}=[R_{23} R_{13}^m \widetilde{\mathcal{A}}_m R_{13}^{m \; \dagger} R_{23}^\dagger]_{\beta\alpha}\;,
\end{equation}
where
\begin{equation}
	\widetilde{\mathcal{A}}_m=\exp[{-i \widetilde{\mathcal{H}}^{(\text{OMSD})}_m L}]
\end{equation}
is the amplitude matrix in the OMSD basis. The neutrino survival/ conversion probabilities can be calculated from,
$P_{\alpha\beta} = |A(\nu_\alpha\to\nu_\beta)|^2$.
They may be written in a compact form in terms of some intermediate quantities. We define
\begin{align}
d_{1}\equiv  \left(1+A-2 i \gamma_{1}^m - C_{13} \right)\frac{\Delta}{L}\;,\quad d_3\equiv \left(1+A-2 i \gamma_{3}^m+C_{13}\right)\frac{\Delta}{L}\;,\quad
\Delta_d\equiv d_3-d_1\;,
\end{align}
and
\begin{align}
D_{1,3}\equiv \left(1+A-i(\gamma_{1}^m+\gamma_{3}^m) \mp \widetilde{C}_{13}^m\right)\frac{\Delta}{L}\;,\qquad \Delta_D\equiv D_3-D_1= 2\widetilde{C}_{13}^m \frac{\Delta}{L}\;,
\end{align}
where 
\begin{equation}
	\widetilde{C}_{13}^m=\sqrt{[C_{13}-i(\gamma_3^m-\gamma_1^m)]^2-(\gamma_{13}^m)^2}\;.
\end{equation}
Applying the Pauli exponentiation technique as discussed in~\cite{Chattopadhyay:2021eba}, we can then directly obtain
\begin{equation}
\widetilde{\mathcal{A}}_{m}=\left(
\begin{array}{ccc}
a \, G_-(L)+G_+(L) &\quad 0 \quad& b \, G_-(L) \\
0 &\quad 1 \quad& 0 \\
b \, G_-(L) &\quad 0 \quad& G_+(L)-a \, G_-(L)\\
\end{array}
\right).
\end{equation}
Here, $G_+(L)$, $G_-(L)$, $a$ and $b$ are given by
\begin{equation}
G_\pm(L)=\frac{1}{2}\left(e^{-i D_3 L}\pm e^{-i D_1 L}\right),\qquad a=-\frac{\Delta_d}{\Delta_D}\;,\qquad b=-2 i \frac{\gamma_{13}^m}{\Delta _D} \frac{\Delta}{L}\;.
\end{equation}
 Finally, the probabilities $P_{\mu\mu}$, $P_{ee}$, $P_{e\mu}$ and $P_{\mu e}$, the ones relevant for most of the neutrino experiments, may be expressed in the OMSD approximation as
\begin{align}
P_{\mu\mu}=&\, \left|c_{23}^2 + s_{23}^2 G_+(t)-s_{23}^2 \left(a \, \cos 2 \theta_{13}^m+ b \, \sin 2 \theta_{13}^m \right) G_-(t)\right|^2,\\
P_{ee}=&\, \left|G_+(t)+\left(a \, \cos 2 \theta_{13}^m+b \,\sin 2 \theta_{13}^m \right) G_-(t)\right|^2, \\
P_{e\mu}=&\, \left|s_{23} \left(b \, \cos 2\theta_{13}^m -a \,\sin 2 \theta_{13}^m \right)G_-(t)\right|^2.
\end{align}
With the OMSD approximation, we have $P_{\mu e}=P_{e \mu}$. Note that,  each of the individual terms --- $G_\pm(t)$, $a$ and $b$ --- are complex in nature.
In this scenario where only $\nu_3$ vacuum mass eigenstate decays, the expressions for probabilities for antineutrinos have the same form as those for neutrinos; however, the numerical values of the parameters $a$, $b$, $\theta_{13}^m$ and $D_{1,3}$ would be different for antineutrinos in the presence of matter. The analytic expansions obtained here will be referred to simply as ``OMSD'' in section~\ref{sec:numcom}, when comparing against the numerically obtained exact results in constant matter density approximation.

The OMSD probabilities obtained above may be explicitly expressed in terms of the quantities $s_{13}$, $\gamma_1^m$, $\gamma_3^m$, and up to $O(\lambda)$ in $\gamma_{13}^m$, using the 2-flavor Zassenhaus expansion employed in~\cite{Chattopadhyay:2021eba}. The relevant expressions are
\begin{align}
		P_{\mu\mu}=&\; \Big(c_{23}^2+s_{23}^2 \Big[(s_{13}^m)^2 e^{-2\gamma_1^m \Delta} +(c_{13}^m)^2 e^{-2\gamma_3^m \Delta} \Big] \Big)^2 -s_{23}^4 \sin^2 2\theta_{13}^m\, e^{-2\gamma_+ \Delta} \sin^2 \Delta_m \nonumber\\
		&- \sin^2 2\theta_{23} \left[(s_{13}^m )^2 e^{-2 \gamma_{1}^m \Delta } \sin^2 \frac{\Delta_-}{2} +(c_{13}^m )^2 e^{-2 \gamma_{3}^m \Delta} \sin^2 \frac{\Delta_+}{2} \right]\nonumber\\
		&+\widetilde{\gamma} s_{23}^2 \sin 2 \theta_{13}^m \Bigg\{ C_{13} \Big[ s_{23}^2 e^{-2\gamma_+ \Delta} \sin 2\Delta_m + c_{23}^2\Big( e^{-2\gamma_{3}^m\Delta} \sin  \Delta_+ -e^{-2\gamma_{1}^m\Delta} \sin \Delta_- \Big) \Big] \Bigg.\nonumber\\
	&\Bigg. \hspace{65pt}+\gamma_- \Bigg[ s_{23}^2 \Big( (c_{13}^m)^2 e^{-4\gamma_3^m\Delta} -(s_{13}^m)^2 e^{-4\gamma_1^m\Delta} -\cos 2\theta_{13}^m e^{-2\gamma_+\Delta} \cos 2 \Delta_m \Big) \Bigg. \nonumber\\
	&\Bigg.  \hspace{95pt} +c_{23}^2 \Big( e^{-2\gamma_{3}^m\Delta} \cos  \Delta_+ -e^{-2\gamma_{1}^m\Delta} \cos \Delta_- \Big) \Bigg] \Bigg\}\;,\\
	P_{ee}=&\, \Big[(c_{13}^m)^2 e^{-2 \gamma_{1}^m \Delta }+ (s_{13}^{m})^2 e^{-2  \gamma _{3}^m \Delta }\Big]^2 - \sin ^2 2 \theta_{13}^m e^{-2 \gamma_+ \Delta} \sin^2 \Delta_m \nonumber \\
	&-\widetilde{\gamma} \sin 2 \theta_{13}^m \Bigg[ C_{13} e^{-2  \gamma_+ \Delta } \sin 2\Delta_m -\gamma_- \Bigg( 2\cos 2 \theta_{13}^m e^{-2 \gamma_+\Delta }\sin^2 \Delta_m \Bigg.\Bigg.\nonumber\\
	&\Bigg. \Bigg.  \hspace{60pt} -\Big[e^{-2 \gamma _{3}^m \Delta }-e^{-2 \gamma_{1}^m \Delta}\Big]  \Big[(s_{13}^m)^2 e^{-2 \gamma _{3}^m \Delta }+ (c_{13}^m)^2 e^{-2 \gamma_{1}^m \Delta}\Big] \Bigg)\Bigg],\\
	P_{e\mu}=&\; s_{23}^2 \sin 2 \theta_{13}^m \Big[  \sin 2 \theta_{13}^m-2 \widetilde{\gamma} \gamma_-   \cos 2 \theta_{13}^m \Big] \left[ \tfrac{1}{4}\left(e^{-2 \gamma_{1}^m \Delta}-e^{-2 \gamma_{3}^m \Delta}\right)^2 + e^{-2 \gamma_+ \Delta}\sin^2 \Delta_m \right]\;.
\end{align}
In the OMSD limit, we have $P_{\mu e}=P_{e\mu}$. Here, $(c_{13}^m)\equiv \cos \theta_{13}^m$ and $(s_{13}^m)\equiv \sin \theta_{13}^m$, and
\begin{equation}
	\Delta_m \equiv C_{13} \Delta\;, \quad \Delta_\pm \equiv \left(1+A\pm C_{13}\right) \Delta\;, \quad \gamma_\pm \equiv \gamma_1^m \pm \gamma_3^m \;, \quad \widetilde{\gamma} \equiv \frac{\gamma_{13}^m}{C_{13}^2+ \gamma_-^2}\;.
\end{equation}
In the no-decay limit, the above equations reduce to the well-known OMSD expressions given in~\cite{Akhmedov:1998xq,Bernabeu:2003yp,Choubey:2003yp,Gandhi:2004bj}.
Note that, the probabilities above are expansions up to $O(\lambda)$ in $\gamma_{13}^m\equiv -\gamma_3 \sin (2\, \delta\theta)$, since $\gamma_3 \sim O(\lambda)$, and $\sin (2\, \delta\theta)$ in principle can be as large as $O(1)$. However, for densities in the crust of the Earth (that are relevant for long-baseline experiments), $\delta \theta=\theta_{13}^m-\theta_{13}$ is very small, and hence higher order corrections due to $\gamma_{13}^m$ are quite small.

The above expressions indicate how, in certain circumstances where the OMSD approximation is valid, the 2-flavor analysis outlined in~\cite{Chattopadhyay:2021eba} allows us to write down compact expressions for neutrino survival/ conversion probabilities in the presence of matter and decay.
These probabilities may be used whenever oscillations due to $\Delta m_{21}^2$ can be ignored.

\section{Developing the 3-flavor Zassenhaus expansion}
\label{sec:zassenhaus}
In~\cite{Chattopadhyay:2021eba}, a resummation procedure for the  Zassenhaus expansion was introduced and applied to calculate the 2-flavor neutrino probabilities in the presence of decay and matter.
In this section, we develop the 3-flavor version of this procedure,
which allows us to calculate the 3-flavor neutrino propagation probabilities with decay in the presence of matter.

The exponential of a sum of two matrices can be decomposed using the inverse Baker-Campbell-Hausdorff (BCH) or Zassenhaus expansion. The expansion may be expressed in terms of an infinite series~\cite{Kimura:2017xxz} as
\begin{equation}{\label{eq:Kimura}}
	e^{\mathbb{X}+\mathbb{Y}}=
	\Big(1+\Big.
	\Big.\sum\limits_{p=1}^\infty \sum\limits_{n_1,...,n_p=1}^\infty \dfrac{n_p...n_1}{n_p (n_p+n_{p-1})...(n_p+...+n_1)} \mathcal{Y}_{n_p}... \mathcal{Y}_{n_1} \Big)e^\mathbb{X}\;,
\end{equation}
where
\begin{equation}
	\label{eq:Yn}
	\mathcal{Y}_{n} =\frac{1}{n!}\mathcal{L}^{n-1}_\mathbb{X} \mathbb{Y}\;.
\end{equation}
Here, $\mathcal{L}$ denotes the commutation operator, i.e.
\begin{equation}
	\mathcal{L}_\mathbb{X} \mathbb{Y}=[\mathbb{X},\mathbb{Y}]\;.
\end{equation}
When truncated up to  first order in $\mathbb{Y}$, only the $p=1$ term contributes, and eq.~(\ref{eq:Kimura}) becomes
\begin{equation}{\label{eq:gamma}}
	e^{\mathbb{X}+\mathbb{Y}} \simeq \Big(1+ \sum\limits_{n_1=1}^\infty \mathcal{Y}_{n_1} \Big)e^\mathbb{X}\;.
\end{equation}
Here the absolute sign($|\cdot|$) denotes the typical scale of a non-zero element of the matrix. 
In order to get closed functional forms for the infinite sums, we need to
find the analytic form of $\mathcal{Y}_n$.

For the general decay scenario in the presence of matter, the effective Hamiltonian in the mass basis in matter (where the Hermitian part of the Hamiltonian is diagonal) takes the form
\begin{equation}
	\mathcal{H}_m=	\frac{\Delta m_{31,\, m}^2}{2 E_\nu} \left(
	\begin{array}{ccc}
		-i \gamma_1^m & -\frac{i }{2} \gamma_{12}^m e^{i \chi_{12}^m} & -\frac{i }{2}\gamma_{13}^m e^{i \chi_{13}^m} \\[5pt]
		-\frac{i }{2} \gamma_{12}^m e^{-i \chi_{12}^m} & \alpha_m -i \gamma_2^m & -\frac{i }{2} \gamma_{23}^m e^{i \chi_{23}^m} \\[5pt]
		-\frac{i }{2} \gamma_{13}^m e^{-i \chi_{13}^m} & -\frac{i }{2} \gamma_{23}^m e^{-i \chi_{23}^m} & 1-i \gamma_3^m \\
	\end{array}
	\right) .
\end{equation}
Note that all the quantities in this section --- $\Delta m_{31,\, m}^2,\, \Delta_m,\, \alpha_m,\,U_m,\, \theta_{ij}^m ,\, \gamma_i^m,\, \gamma_{ij}^m$ and $\chi_{ij}^m$ --- are to be taken in the presence of matter. These expressions are also applicable for vacuum, i.e. at vanishing matter density. 

In order to calculate the amplitude matrix $\mathcal{A}_m=e^{-i \mathcal{H}_m L}$ in the matter basis, we define the matrix $\mathbb{A}\equiv -i \mathcal{H}_m L$. This matrix can be decomposed into $\mathbb{A}=\mathbb{X}+\mathbb{Y}$, where $\mathbb{X}$ is a diagonal matrix, and $\mathbb{Y}$ consists of the off-diagonal contributions due to the $\gamma_{ij}$ terms:
\begin{equation}
	\mathbb{X}= -2 i \left(
	\begin{array}{ccc}
		-i \gamma_1^m& 0& 0 \\[5pt]
		0& \alpha_m-i\gamma_2^m & 0\\[5pt]
		0 & 0& 1-i\gamma_3^m \\
	\end{array}
	\right)\Delta_m\;, \quad	\mathbb{Y}= - \left(
	\begin{array}{ccc}
		0&  \gamma_{12}^m e^{i \chi_{12}^m} &  \gamma_{13}^m e^{i \chi_{13}^m } \\[5pt]
		\gamma_{12}^m e^{-i \chi_{12}^m} &0 &  \gamma_{23}^m e^{i \chi_{23}^m } \\[5pt]
		\gamma_{13}^m e^{-i \chi_{13}^m} & \gamma_{23}^m e^{-i \chi_{23}^m} &0 \\
	\end{array}
	\right)\Delta_m\;.
\end{equation}
Note that the matrix $\mathbb{Y}$ can be further decomposed into
\begin{equation}
	\mathbb{Y}=\mathbb{Y}_{12}+\mathbb{Y}_{13}+\mathbb{Y}_{23}\;,
\end{equation}
where $\mathbb{Y}_{12}$, $\mathbb{Y}_{13}$ and $\mathbb{Y}_{23}$ are hermitian matrices with the only non-zero elements
\begin{equation}
	[\mathbb{Y}_{ij}]_{ij}= -\gamma_{ij}^m e^{i\chi_{ij}^m} \cdot \Delta_m\;,\qquad 	[\mathbb{Y}_{ij}]_{ji}= -\gamma_{ij}^m e^{-i\chi_{ij}^m} \cdot \Delta_m \;.
\end{equation}
Using $\mathbb{Y}_{12}$, $\mathbb{Y}_{13}$ and $\mathbb{Y}_{23}$, we can calculate $\mathcal{Y}_n$ in eq.~(\ref{eq:Yn}) as
\begin{align}
	\mathcal{Y}_n=\frac{1}{n!}&\left( \Delta_{12}^{n-1} \Sigma_{12}^{n-1} \mathbb{Y}_{12} +\Delta_{13}^{n-1} \Sigma_{13}^{n-1} \mathbb{Y}_{13} + \Delta_{23}^{n-1} \Sigma_{23}^{n-1} \mathbb{Y}_{23} \right),
\end{align}
where we have defined $\Sigma_{12}=\text{diag}[(1,-1,0)]$,  $\Sigma_{13}=\text{diag}[(1,0,-1)]$, $\Sigma_{23}=\text{diag}[(0,1,-1)]$, and
\begin{equation}
	\Delta_{ij} \equiv \mathbb{X}_{ii}-\mathbb{X}_{jj}\;.
\end{equation}
The amplitude matrix in matter basis  calculated up to first order in $\gamma_{ij}$ can be expressed as
\begin{equation}
\mathcal{A}_m=\left(
	\begin{array}{ccc}
		e^{-2 \gamma_1^m \Delta_m} & \quad 2 \Delta_m \dfrac{\gamma_{12}^m \, e^{i\chi_{12}^m}}{\Delta_{12}} g_{12}(L) \quad & 2 \Delta_m \dfrac{\gamma_{13}^m \, e^{i\chi_{13}^m}}{\Delta_{13}} g_{13}(L)  \\[7pt]
		 2 \Delta_m \dfrac{\gamma_{12}^m\, e^{-i\chi_{12}^m}}{\Delta_{12}} g_{12}(L) & \quad e^{-2 i \alpha_m \Delta_m -2\gamma_2^m \Delta_m} \quad & 2 \Delta_m \dfrac{\gamma_{23}^m \, e^{i\chi_{23}^m}}{\Delta_{23}} g_{23}(L)\\[7pt]
		 2 \Delta_m \dfrac{\gamma_{13}^m e^{-i\chi_{13}^m}}{\Delta_{13}} g_{13}(L) & \quad  2 \Delta_m \dfrac{\gamma_{23}^m e^{-i\chi_{23}^m}}{\Delta_{23}} g_{23}(L) \quad & e^{-2 i \Delta_m -2\gamma_3^m \Delta_m } \\[7pt]
	\end{array}
	\right),
	\label{eq:AmZassenhaus}
\end{equation}
where $g_{ij}(L) \equiv \frac{1}{2}\left(\exp[{\mathbb{X}_{jj}}]-\exp[{\mathbb{X}_{ii}}]\right)$. The probability in flavor basis can then be calculated via
\begin{equation}
	P_{\alpha\beta}=\Big| [U_m \mathcal{A}_m U_m^\dagger]_{\beta \alpha} \Big|^2\;.
\end{equation}
We now present the probabilities in vacuum and in the presence of matter.
\subsection{Probabilities in vacuum, expanded in $\alpha$, $s_{13}$, and $\gamma_3$}
\label{sec:zassenhaus1}
We use the matter amplitude in eq.~(\ref{eq:AmZassenhaus}) to calculate the probabilities, we expand perturbatively in term of the parameters, $\alpha$, $s_{13}$ and $\gamma_3$. 
Note that the expressions for the amplitude is valid also in vacuum, just by dropping the suffix `m'.

The normalized decay widths $\gamma_i$'s and $\gamma_{ij}$'s, defined in the vacuum basis can be severely constrained for a neutrino propagating through vacuum.
This gives, effectively, $\gamma_{1},\;\gamma_2,\; \gamma_{12},\; \gamma_{13},\; \gamma_{23} \to 0$, so we may work in the limit of only $\nu_3$  decay, with $\gamma_3 \sim O(\lambda)$.
The survival/ conversion probability may be expressed as a summation of 2 individual terms in this case:
\begin{equation}
	P_{\alpha\beta}=P^{(0)}_{\alpha\beta}+P^{(\gamma_3)}_{\alpha\beta}\;.
\end{equation}
Here $P^{(0)}_{\alpha\beta}$ is the no decay contribution, while $P^{(\gamma_3)}_{\alpha\beta}$ is the contribution from $\gamma_3$.
The neutrino survival/ conversion probabilities $P_{\mu\mu}$, $P_{ee}$, $P_{e\mu}$ and $P_{\mu e}$ are as given below. Note that due to the largeness of $\theta_{23}$, the effect of $\nu_3$ decay is the most prominent in $P_{\mu\mu}$:

\begin{align}
	P^{(0)}_{\mu\mu}=&\, 1- \sin ^2 2\theta _{23} \sin ^2\Delta\nonumber\\
	& +4  s_{13}^2 s_{23}^2\cos 2 \theta _{23} \sin^2 \Delta + \alpha \sin^2 2\theta_{23}\, c_{12}^2\; \Delta  \sin 2 \Delta+O(\lambda^3)\;,\label{414}\\
	P^{(\gamma_3)}_{\mu\mu}=& -\gamma _3 \Delta  \left(\sin ^2 2 \theta _{23}\cos 2 \Delta +4 s_{23}^4\right)\nonumber\\
	& + \gamma _3^2 \Delta ^2 \left(\sin ^2 2\theta _{23} \cos 2 \Delta +8 s_{23}^4\right)+O(\lambda^3)\;,
\end{align}
The probabilities $P_{ee}$, $P_{e\mu}$ and $P_{\mu e}$ are calculated up to $O(\lambda^3)$ since the modifications due to $\gamma_3$ only manifest at the third order:
\begin{align}
	P_{ee}^{(0)}=&\, 1-4 s^2_{13} \sin ^2\Delta +O(\lambda^4)\;, \\
	P_{ee}^{(\gamma_3)}=&-4\, \gamma _3 \, s^2_{13}\, \Delta  \cos 2 \Delta +O(\lambda^4)\;, \\
	P^{(0)}_{e\mu}=&\, 4 s_{13}^2 s_{23}^2 \sin^2 \Delta +2 \alpha \, s_{13} \, \sin 2 \theta _{12} \sin 2 \theta _{23} \; \Delta \, \cos \left(\Delta -\delta _{\text{CP}}\right)  \sin \Delta +O(\lambda^4)\;, \\
	P^{(\gamma_3)}_{e\mu}=&-8 \gamma _3 \, s_{13}^2  s_{23}^2 \; \Delta  \sin ^2\Delta+O(\lambda^4)\;. \label{419}
\end{align}
Since the above equations are for neutrino propagation in vacuum, $P_{\mu e}$ is given by the replacement rule
\begin{align}
	P_{\mu e} \equiv &\; P_{e\mu} (\delta_{\text{CP}} \to -\delta_{\text{CP}})\;.
\end{align}
The antineutrino propagation probabilities are also obtained by the replacement rule
\begin{equation}
	P_{\bar{\alpha}\bar{\beta}}=P_{\alpha\beta} (\delta_{\text{CP}} \to -\delta_{\text{CP}})\;.
\end{equation}

\subsection{Probabilities in matter, expanded in $\alpha_m$, $s^m_{13}$, $\gamma^m_{i}$'s and $\gamma_{ij}^m$'s}

In matter, the stringent decay constraints on the general decay matrix may not survive and all the decay elements may be non-zero.
However, note that the decay elements $\gamma_i^m$'s and $\gamma_{ij}^m$'s are defined in the matter basis (where the Hermitian part of the Hamiltonian is diagonal).
In the presence of large matter densities, the rotation induced by matter effects to the decay matrix may be large, and we may even have $\gamma_{ij} \sim O(\lambda)$. The decay matrix $\Gamma$ defined in vacuum, with its terms represented by $\gamma_i$ and $\gamma_{ij}$, can be connected with the decay matrix $\Gamma_m$ defined in the matter basis (with elements $\gamma_i^m$ and $\gamma_{ij}^m$) via
\begin{equation}
\Gamma_m= U_m^\dagger\; U \; \Gamma \; U^\dagger\; U_m \;.
\end{equation}
Recall that, due to the mismatch between the vacuum and the matter eigenstates, all $\gamma_i^m$'s and $\gamma_{ij}^m$'s would be non-zero even if only the vacuum mass eigenstate $\nu_3$ were to decay.
Let us express the full probability in terms of
\begin{equation}
	P_{\alpha\beta}=P^{(0)}_{\alpha\beta}+P^{(\gamma_3)}_{\alpha\beta}+P^{(\Gamma)}_{\alpha\beta}\;,
\end{equation}
 where the new $P^{(\Gamma)}_{\alpha\beta}$ term denotes the contribution of the general decay matrix apart from that of the $\gamma_3^m$.
The first two terms may be obtained by applying the replacement rules 
\begin{equation}
	\theta_{ij}\to\theta_{ij}^m ,\quad \gamma_{3} \to \gamma_{3}^m,\quad \chi_{ij} \to \chi_{ij}^m,\quad \Delta \to \Delta_m,\quad \alpha \to \alpha_m 
\end{equation}
to eq.~(\ref{414})-(\ref{419}), i.e. the suffix `$m$' should be reintroduced to represent that we are working in matter basis.
The additional contributions from the general decay matrix may be given as
\begin{align}
	P^{(\Gamma)}_{\mu\mu}=&\, \sin 2 \theta^m_{23} \left(\gamma^m_{13} s^m_{12} \cos \chi^m_{13}-\gamma^m_{23} c^m_{12} \cos \chi^m_{23} \right)\sin 2 \Delta_m \;,\\
	P^{(\Gamma)}_{ee}=&-4 \gamma_1^m \Delta_m (c^m_{12})^2 -4 \gamma_2^m \Delta_m (s^m_{12})^2-2 \gamma_{12}^m \Delta_m  \sin 2 \theta^m_{12} \cos \chi^m_{12} \nonumber\\
	&-2 s^m_{13} \left[\gamma^m_{13} c^m_{12} \cos \left(\delta_{\text{CP}} + \chi^m_{13} \right) +\gamma^m_{23} s^m_{12} \cos \left(\delta_{\text{CP}}+\chi^m_{23}\right)\right] \sin 2 \Delta_m \;,\\
	P^{(\Gamma)}_{e\mu}=&-4 s^m_{13} (s^m_{23})^2 \left[\gamma^m_{13} c^m_{12} \sin \left(\delta_{\text{CP}}+\chi^m_{13}\right)+\gamma^m_{23} s^m_{12} \sin \left(\delta _{\text{CP}}+\chi^m_{23}\right)\right]\sin^2\Delta_m \;.
\end{align}
We also have
\begin{align}
	P_{\mu e}^{(\Gamma)} = &\, P_{e\mu}^{(\Gamma)} (\delta_{\text{CP}}^m \to -\delta_{\text{CP}}^m, \chi_{ij}^m \to -\chi_{ij}^m)\;.
\end{align}
The probabilities associated with antineutrino propagation are given by
\begin{equation}
	P_{\bar{\alpha}\bar{\beta}}=P_{\alpha\beta} (\delta_{\text{CP}}^m \to -\delta_{\text{CP}}^m, \chi_{ij}^m \to -\chi_{ij}^m)\;,
\end{equation}
with all parameters in matter calculated using $A\to -A$.

Since the amplitude derived above is only accurate up to $O(\gamma^m_{ij})$, our probability expression will only be correct up to $O(\lambda)$, even when $s_{13}^m$ is not too large. For matter densities in the Earth crust, the mismatch between matter and vacuum basis is small, and our expressions should be correct up to higher orders in $\lambda$.

In order to get analytic expressions that are valid over a wide range of matter densities, we need explicit dependence on the matter potential. In the next 2 sections, applying the Cayley-Hamilton theorem, we derive the neutrino survival/ conversion probabilities with complete dependence on the matter potential term, first for the decay of $\nu_3$ only, and afterwards for the general decay matrix $\Gamma$.

\section{Explicit matter dependence, with decay of $\nu_3$ only}
\label{sec:CayleyHamiltonnu3}
In this section we will obtain compact analytic forms for the relevant neutrino probabilities with explicit matter dependence.
We use the effective Hamiltonian $\mathcal{H}_f^{(\gamma_3)}$ in eq.~(\ref{eq:Hnu3decay}), denoting the case when only the vacuum mass eigenstate $\nu_3$ decays. We employ the Cayley-Hamilton theorem~\cite{Ohlsson:1999xb} to calculate relevant probabilities with explicit dependence on the `normalized' matter potential $A\equiv 2 E_\nu V_{cc}/\Delta m^2_{31}$.

Using the Cayley-Hamilton theorem, any function $g(\mathbb{M})$ of a matrix $\mathbb{M}$ can be expressed in terms of its eigenvalues as
\begin{equation}
	g(\mathbb{M})=\sum_{i=1}^k M_i\; g\left(\lambda _i\right)\; ,\quad \text{ with }\quad M_i \equiv \prod_{j=1,j\neq i}^{k} \frac{1}{\lambda_i-\lambda_j}(\mathbb{M}-\lambda_j \mathbb{I})\;,
\end{equation}
where $\lambda_i$'s are the distinct eigenvalues of the matrix $\mathbb{M}$.
Taking $\mathbb{M}=-i \mathcal{H}_f^{(\gamma_3)} L$, the probability amplitude matrix $\mathcal{A}_f$, such that $[\mathcal{A}_f]_{\beta\alpha}=A(\nu_\alpha \to \nu_\beta)$, can be expressed as

\begin{equation}
	\begin{split}
		\mathcal{A}_f=\, \exp[{-i \mathcal{H}_f^{(\gamma_3)} L}]=&\; \frac{e^{-i E_1 L}}{(E_1-E_2) (E_1-E_3)}  \Big[\mathcal{H}_f^{(\gamma_3)}-E_2 \mathbb{I}\Big]\Big[\mathcal{H}_f^{(\gamma_3)}-E_3 \mathbb{I}\Big] \\
		&+\frac{e^{-i E_2 L}}{(E_2-\text{E}_1) (E_2-E_3)} \Big[\mathcal{H}_f^{(\gamma_3)}-E_1 \mathbb{I}\Big]\Big[\mathcal{H}_f^{(\gamma_3)}-E_3 \mathbb{I}\Big]\\
		&+\frac{e^{-i E_3 L}}{(E_3-E_1) (E_3-E_2)} \Big[\mathcal{H}_f^{(\gamma_3)}-E_1 \mathbb{I}\Big]\Big[\mathcal{H}_f^{(\gamma_3)}-E_2 \mathbb{I}\Big]\;.
	\end{split}
\end{equation}
Here, $E_1,\; E_2$ and $E_3$ are the eigenvalues of the effective Hamiltonian. The survival/ conversion probabilities can be calculated as $P_{\alpha\beta}=|A(\nu_\alpha \to \nu_\beta)|^2$.
In the next two subsections, we employ perturbative expansions in terms of different choices of small parameters that may be valid for different parameter regimes.
Even for the next generation of neutrino experiments, sensitivity beyond a few percent is not expected to be achieved. Therefore, perturbative expansions up to $O(\lambda^2)-O(\lambda^3)$ for the probabilities should suffice.

\subsection{Probabilities expanded in $s_{13}$, $\alpha$ and $\gamma_3$}
\label{sec:fullexpansion}
The analytic expansions in $s_{13}$, $\alpha$ and $\gamma$, up to $O(\lambda^2)$ or $O(\lambda^3)$, are especially useful because of their simple compact form. 
We take $s_{13},\gamma_3 \sim O(\lambda)$, $\alpha \sim O(\lambda^2)$.
The eigenvalues of the Hamiltonian may be expressed in the form $E_i = E_i^{(0)}+E_i^{(\gamma_3)}$, where $E_i^{(0)}$ is the no-decay contribution to the eigenvalues, and $E_i^{(\gamma_3)}$ is the modification due to the decay of $\nu_3$ mass eigenstate in vacuum. These can be given as
\begin{align}
		E_1^{(0)}+E_1^{(\gamma_3)} \equiv E_1=&\; \frac{\Delta m_{31}^2}{2 E_\nu}\left(A+\alpha \,  s_{12}^2+ s_{13}^2 \frac{A }{A -1}-i \gamma_3 \, s_{13}^2 \frac{A^2}{(A-1)^2}\right)+O(\lambda^4)\;,\\
		E_2^{(0)}+E_2^{(\gamma_3)} \equiv E_2=&\; \frac{\Delta m_{31}^2}{2 E_\nu}\Big(\alpha \, c_{12}^2\Big)+O(\lambda^4)\;,\\
		E_3^{(0)}+E_3^{(\gamma_3)} \equiv E_3=&\; \frac{\Delta m_{31}^2}{2 E_\nu} \left(1-i \gamma _3- s_{13}^2 \frac{A}{A-1}+i \gamma _3 \, s_{13}^2 \frac{ A^2}{(A-1)^2}\right)+O(\lambda^4)\;.
\end{align}
It may be observed that $E_3$ gets modified at $O(\lambda)$, $E_1$ has a $O(\lambda^3)$ modification, while $E_2$ has no modifications due to decay up to $O(\lambda^3)$.

Following our earlier convention, we express the probabilities as $P_{\alpha\beta}=P^{(0)}_{\alpha\beta}+P^{(\gamma_3)}_{\alpha\beta}$, where $P^{(0)}_{\alpha\beta}$ denotes the no-decay contribution and $P^{(\gamma_3)}_{\alpha\beta}$ denotes the modifications due to the decay of $\nu_3$.
We present the expressions for $P_{\mu\mu}$, $P_{ee}$, $P_{e\mu}$, and $P_{\mu e}$, relevant for neutrino experiments.
The largest effect of $\gamma_3$ is expected in $P_{\mu\mu}$, where we get
\begin{align}
	P^{(0)}_{\mu\mu}=&\, 1-\sin ^2 2 \theta _{23} \sin ^2\Delta \nonumber\\
	&-\frac{2}{A-1} s_{13}^2 \sin ^2 2\theta _{23} \left(\sin\Delta \cos A\Delta \, \frac{\sin[ (A-1) \Delta]}{A-1}-\frac{A}{2}\, \Delta \sin 2\Delta\right)\nonumber\\
	& -4 s_{13}^2 s_{23}^2 \frac{\sin ^2[(A-1) \Delta ]}{(A-1)^2}+\alpha \, c_{12}^2 \, \sin^2 2\theta_{23} \; \Delta \sin 2\Delta +O(\lambda^3)\;, \label{eq:expansionmumu0} \\
	P^{(\gamma_3)}_{\mu\mu}=& -\gamma _3 \Delta  \left(\sin ^2 2 \theta _{23}\cos 2 \Delta +4 s_{23}^4\right)\nonumber\\
	& + \gamma _3^2 \Delta ^2 \left(\sin ^2 2\theta _{23} \cos 2 \Delta +8 s_{23}^4\right)+O(\lambda^3)\;.
\end{align}
For $P_{ee}$, $P_{e \mu}$, and $P_{\mu e}$, the terms  with $\gamma_3$ appear at  $O(\lambda^3)$. Therefore, we present these probabilities up to $O(\lambda^3)$:
 \begin{align}
 	P^{(0)}_{ee}=&\, 1-4 s_{13}^2 \frac{\sin ^2 [(A-1)\Delta]}{(A-1)^2}+O(\lambda^4) \;, \\
 	P^{(\gamma_3)}_{ee}=&\, \gamma _3 \, s_{13}^2 \left(4 A\frac{\sin [2 (A-1) \Delta] }{(A-1)^3}-4 \Delta \frac{1+ A^2}{(A-1)^2}+8\Delta \frac{\sin^2 [(A-1)\Delta]}{(A-1)^2}\right)+O(\lambda^4) \;, \\
 	P^{(0)}_{e\mu}=&\, 4 s_{13}^2 s_{23}^2 \frac{\sin ^2[(A-1) \Delta]}{(A-1)^2}\nonumber\\
 	&+2 \alpha \, s_{13} \, \sin 2 \theta _{12} \sin 2 \theta _{23} \cos \left(\Delta -\delta _{\text{CP}}\right)\frac{\sin [(A-1) \Delta ]}{A-1}\frac{\sin A \Delta}{A}+O(\lambda^4) \; ,\\
 	P^{(\gamma_3)}_{e\mu}=&-8 \gamma _3 \, s_{13}^2 s_{23}^2 \; \Delta \, \frac{\sin ^2[(A-1) \Delta]}{(A-1)^2}+O(\lambda^4)\;. \label{eq:expansionmumugamma3}
 \end{align}
Note that $P_{\mu e} = \, P_{e \mu} (\delta_{\text{CP}} \to -\delta_{\text{CP}})$. The  antineutrino oscillation probabilities are given by
$ P_{\bar{\alpha}\bar{\beta}}=P_{\alpha\beta} (\delta_{\text{CP}} \to -\delta_{\text{CP}}\, , \; A\to -A)\;. $
The probabilities obtained here will be referred to as ``Full-expansion'' in section~\ref{sec:numcom}.
In the vacuum limit, i.e. in the limit of  $A \to 0$, the survival/ conversion probabilities given above match the results derived via Zassenhaus expansion in section~\ref{sec:zassenhaus}. Moreover, the expressions for $P_{\alpha\beta}^{(0)}$ above match those given in~\cite{Akhmedov:2004ny} to appropriate orders, as expected.

The perturbative expansions in eq.~(\ref{eq:expansionmumu0})-(\ref{eq:expansionmumugamma3}) are valid as long as $\alpha \Delta \lesssim 1$ and $\gamma_3 \Delta \lesssim 1$.
For example, for the first oscillation peak of both T2K and DUNE, we find that our perturbative analysis is valid.
Since $\gamma_3 \sim O(\lambda)$, when $\lambda \Delta \sim 1$, then our perturbative expansion is expected to deviate from the numerically obtained values. This would happen at large values of $L/E$, i.e. at longer baselines, and/or lower energies.
With an exact dependence on $\gamma_3$, we would be able to increase the region of validity of our analytic expressions to the point where $\alpha \Delta \sim \lambda^2 \Delta \lesssim 1$.
This would be done in the next subsection.

\subsection{Probabilities expanded in $s_{13}$, and $\alpha$; exact in $\gamma_3$}
\label{sec:expansion}

An expansion that retains the exact dependence on $\gamma_3$ is expected to give further insight into the probabilities, demonstrating the $e^{-\gamma_3 \Delta}$ behavior explicitly. As discussed above, such an expansion will also be applicable for higher values of $\Delta$, and hence to lower energy regimes for a given baseline.

Taking $s_{13}\sim O(\lambda)$ and $\alpha\sim O(\lambda^2)$,
the eigenvalues of the Hamiltonian are
\begin{align}
	E_1 \simeq &\; \frac{\Delta m_{31}^2}{2 E_\nu} \left(A+\alpha \, s_{12}^2+s_{13}^2 \frac{A \left(1-i \gamma _3\right)}{A-(1-i \gamma _3)}\right)+O(\lambda^4)\;,\\
	E_2 \simeq &\; \frac{\Delta m_{31}^2}{2 E_\nu}\Big(\alpha \, c_{12}^2\Big)+O(\lambda^4)\;,\\
	E_3 \simeq &\; \frac{\Delta m_{31}^2}{2 E_\nu} \left(1-i \gamma _3-s_{13}^2 \frac{A \left(1-i \gamma _3\right)}{A-(1-i \gamma _3)}\right)+O(\lambda^4)\;.
\end{align}
Employing the Cayley-Hamilton procedure as in Section 5.1, the neutrino survival probability $P_{\mu\mu}$ becomes
\begin{align}
	P_{\mu\mu}=&\; \Bigg|c_{23}^2+s_{23}^2 e^{-2i\left(1-i\gamma_3 \right) \Delta} -2 i \alpha \, c_{12}^2 c_{23}^2 \, \Delta +s_{13}^2 s_{23}^2 \Bigg(e^{-2 i A \Delta }\frac{\left(1-i\gamma _3\right)^2}{\left[A-(1-i \gamma _3)\right]^2} \Bigg. \Bigg. \nonumber\\
	&\Bigg.\Bigg.\; +e^{-2i\left(1-i\gamma_3 \right) \Delta} \Big[2iA\Delta\left[A-(1-i\gamma_3)\right]-(1-i\gamma_3)\Big]\frac{1-i\gamma _3}{\left[A-(1-i \gamma _3)\right]^2}\Bigg) \Bigg|^2 +O(\lambda^3)\;.
	\label{eq:PmumuExpansion}
\end{align}
Note that the leading contribution of the decay parameter $\gamma_3$ is through the $s_{23}^2$ term. The contribution from the first two terms in the amplitude above is
\begin{align}
	P_{\mu\mu}^\text{leading}=&\, c_{23}^4  + s_{23}^4 \, e^{-4\gamma_3 \Delta} +  2 s_{23}^2 c_{23}^2  \cos (2 \Delta) e^{-2\gamma_3 \Delta}\nonumber\\
	=&\, 1 - \sin^2 2\theta_{23} \sin^2 \Delta -s_{23}^4 \left(1-e^{-4\gamma_3 \Delta}\right) - 2 s_{23}^2 c_{23}^2  \cos (2 \Delta)\left(1-e^{-2\gamma_3 \Delta}\right)\;. \label{eq:pmumuleading}
\end{align}
If the value of $\gamma_3$ is indeed $\sim O(\lambda)$, the above terms would lead to significant deviations from the standard 3-neutrino oscillation probabilities, and may give the first indications of decaying $\nu_3$.

The survival probability $P_{ee}$ and the conversion probabilities $P_{e \mu}$  and $P_{\mu e}$ are
\begin{align}
	P_{ee}=&\,1-2 s_{13}^2 \Bigg[ \Big(1-e^{-2 \gamma_3 \Delta } \cos [2 (A-1) \Delta] \Big)\left[\frac{1+\gamma _3^2}{(A-1)^2+\gamma _3^2}-\frac{2 A^2 \gamma _3^2}{\left[(A-1)^2+\gamma _3^2\right]^2}\right] \Bigg.\nonumber\\
	&\Bigg. +e^{-2 \gamma _3 \Delta } \frac{2 A \gamma _3 \left(1-A+\gamma _3^2\right)}{\left[(A-1)^2+\gamma _3^2\right]^2}\sin [2 (A-1) \Delta] + \frac{2 A^2 \gamma _3 \Delta }{(A-1)^2+\gamma _3^2}\Bigg] +O(\lambda^4)\;, \label{eq:PeeExpansion} \\
	P_{ e \mu }=&\, s_{13}^2 s_{23}^2 \Big(1+e^{-4 \gamma _3 \Delta }-2 e^{-2 \gamma _3 \Delta } \cos [2 (A-1) \Delta] \Big)\frac{\gamma _3^2+1}{(A-1)^2+\gamma _3^2}\nonumber\\
	&+\alpha \, s_{13} \, \sin 2 \theta _{12} \sin 2 \theta _{23} \frac{\sin A \Delta}{A} \nonumber\\
	&\times \Bigg[\Big( \sin \left[(A-2) \Delta +\delta _{\text{CP}} \right]e^{-2 \gamma _3 \Delta }+\sin \left[A \Delta -\delta _{\text{CP}}\right]\Big)\frac{(A-1)-\gamma _3^2}{(A-1)^2+\gamma _3^2}\Bigg.\nonumber\\
	&\Bigg.+\gamma _3\Big(\cos \left[A \Delta -\delta _{\text{CP}}\right]- \cos \left[(A-2) \Delta +\delta _{\text{CP}}\right] e^{-2 \gamma _3 \Delta }\Big)\frac{A}{(A-1)^2+\gamma _3^2}\Bigg]+O(\lambda^4)\;. \label{eq:PemuExpansion}
\end{align}
With $P_{\mu e}=P_{e\mu}(\delta_{\text{CP}} \to - \delta_{\text{CP}})$, and
	$ P_{\bar{\alpha}\bar{\beta}}=P_{\alpha\beta} (\delta_{\text{CP}} \to -\delta_{\text{CP}}\, , \; A\to -A)$
. The probabilities obtained here will be referred to as ``Expansion'' in section~\ref{sec:numcom}. Since both the ``Expansion'' and the ``Full-expansion'' give the same results in the absence of decay, we shall refer to these simply as ``Expansion'' while discussing the probabilities without decay. Note that in eq.~(\ref{eq:PmumuExpansion})-(\ref{eq:PemuExpansion}) match~\cite{Akhmedov:2004ny} in the limit $\gamma_3\to 0$, as expected. However, they bring out the complex nature of dependence of probabilities on $\gamma_3$.

Since we have an exact dependence on $\gamma_3$, our expansion is valid for all values of $\Delta$ such that $\alpha\Delta\lesssim 1$, i.e. for $\lambda^2 \Delta \lesssim 1$. Therefore, the analytic expression with expansion in $s_{13}$ and $\alpha$ and an exact dependence on $\gamma_3$ and the matter potential $A$ is the most suitable for longer baselines or lower energies.

\section{Explicit matter dependence, with a general decay matrix $\Gamma$}
\label{sec:CayleyHamiltonGamma}
As argued in section~\ref{sec:hamiltonian_gen}, when we allow maximum possible values for
the elements of the decay matrix $\Gamma$ , we have
\begin{equation}
 \gamma_{1}\, , \gamma_{2} \sim O(\lambda^3)\;, \qquad \gamma_3 \sim O(\lambda)\;,\qquad \gamma_{12} \sim O(\lambda^3)\;,\qquad \gamma_{13},\, \gamma_{23} \sim O(\lambda^2)\;.
\end{equation}
Note that these elements are defined in the vacuum basis.
\subsection{Probabilities expanded in $s_{13}$, $\alpha$, $\gamma_i$ and $\gamma_{ij}$}
The eigenvalues of the effective Hamiltonian $\mathcal{H}_f$ in eq.~(\ref{eq:Hgendecay}) due to the inclusion of the general decay matrix $\Gamma$ can be  expressed as
\begin{equation}
	E_i =E_i^{(0)} +E_i^{(\gamma_3)}+E_i^{(\Gamma)}\;.
\end{equation}
While $E_i^{(0)}$ and $E_i^{(\gamma_3)}$ are given in section~\ref{sec:fullexpansion}, the additional terms $E_i^{(\Gamma)}$ are
\begin{align}
E_1^{(\Gamma)}=\frac{\Delta m_{31}^2 }{2 E_\nu}& \Bigg[-i \gamma _1 \, c_{12}^2-i \gamma _2 \, s_{12}^2-i \gamma_{12} \, s_{12} c_{12} \cos \chi _{12} \Bigg. \\
&\Bigg.-i s_{13} \Big(\gamma _{13} \, c_{12} \cos \left[\delta _{\text{CP}}+\chi _{13}\right]+\gamma _{23} \, s_{12} \cos \left[\delta _{\text{CP}}+\chi _{23}\right]\Big) \frac{A }{A-1} \Bigg]+O(\lambda^4) \;,\nonumber \\
E_2^{(\Gamma)}=\frac{\Delta m_{31}^2 }{2 E_\nu}&\Bigg[-i \gamma _1\, s_{12}^2-i\gamma _2 \, c_{12}^2-i \gamma _{12} \, s_{12} c_{12} \, \cos \chi _{12} \Bigg]+O(\lambda^4)\;,\\
E_3^{(\Gamma)}=\frac{\Delta m_{31}^2 }{2 E_\nu}&\Bigg[i \, s_{13} \Big(\gamma _{13} \, c_{12} \cos \left[\delta _{\text{CP}}+\chi _{13}\right]+\gamma _{23}\, s_{12} \cos \left[\delta _{\text{CP}}+\chi _{23}\right]\Big)\frac{A}{A-1}\Bigg]+O(\lambda^4)\;.
\end{align}
Applying Cayley-Hamilton theorem to the effective Hamiltonian $\mathcal{H}_f^{(\Gamma)}$, as defined in eq.~(\ref{eq:Hgendecay}), gives the neutrino probabilities
\begin{equation}
P_{\alpha\beta}=P^{(0)}_{\alpha\beta}+P^{(\gamma_3)}_{\alpha\beta}+P^{(\Gamma)}_{\alpha\beta}\;,
\end{equation}
where $P^{(\Gamma)}_{\alpha\beta}$ is the extra contribution to the probability corresponding due to the new diagonal elements $\gamma_1,\, \gamma_2$ and off-diagonal elements $\gamma_{12}$, $\gamma_{13}$ and $\gamma_{23}$. The expressions for the $P^{(0)}_{\alpha\beta}$ and $P^{(\gamma_3)}_{\alpha\beta}$ are already given in section~\ref{sec:fullexpansion}.
The $P_{\alpha\beta}^{(\Gamma)}$ probabilities are given as
\begin{align}
	P^{(\Gamma)}_{\mu\mu}=&\, \sin 2 \theta _{23} \left(\gamma _{13} \, s_{12} \cos \chi_{13}-\gamma _{23} \, c_{12} \cos \chi _{23}\right)\sin 2 \Delta +O(\lambda^3)\;,\\
	P^{(\Gamma)}_{ee}=&-4 \gamma _1 \, c_{12}^2 \Delta -4 \gamma _2\, s_{12}^2 \Delta  -2 \gamma _{12} \, \Delta  \sin 2 \theta _{12} \cos \chi _{12}\nonumber\\
	&+2 s_{13} \Big(\gamma _{13} \, c_{12} \cos \left[\delta _{\text{CP}}+\chi _{13}\right]+\gamma _{23}  \, s_{12} \cos \left[\delta _{\text{CP}}+\chi _{23}\right]\Big)\left(\frac{\sin [2 (A-1) \Delta] }{(A-1)^2}-\frac{2 A \Delta }{A-1}\right)\nonumber\\
	&+O(\lambda^4)\; ,\\
	P^{(\Gamma)}_{e\mu}=&-4 s_{13}  s_{23}^2 \left(\gamma _{23} \, s_{12} \sin \left[\delta _{\text{CP}}+\chi _{23}\right]+\gamma _{13} \, c_{12} \sin \left[\delta _{\text{CP}}+\chi _{13}\right]\right)\frac{\sin ^2 [(A-1) \Delta]  }{(A-1)^2}+O(\lambda^4)\;,
\end{align}
and $P_{\mu e}=P_{e \mu}(\delta_{\text{CP}} \to -\delta_{\text{CP}}, \chi_{ij} \to -\chi_{ij})$. The probabilities for antineutrino are given by
	\begin{equation}
		P_{\bar{\alpha}\bar{\beta}}=P_{\alpha\beta} (\delta_{\text{CP}} \to -\delta_{\text{CP}}\, ,\; \chi_{ij} \to -\chi_{ij}\, , \; A \to -A)\;.
	\end{equation}

From the above expressions, we make the following observations:
 \begin{itemize}
 	\item The $P^{(\Gamma)}_{\mu\mu}$ contributions do not have any matter dependence, similar to the observation made in section~\ref{sec:expansion}.
 	
 	\item The modifications to $P_{\mu\mu}$ due to the off-diagonal decay components are $O(\lambda^2)$, which are subleading as compared to the contributions of $\gamma_3$.
 	
 	\item The probability modifications $P_{ee}^{(\Gamma)}$, $P_{e\mu}^{(\Gamma)}$ are $\sim O(\lambda^3)$. Therefore, the effects of neutrino decay would continue to be hard to observe in the $P_{ee}$, $P_{e \mu}$, and $P_{\mu e}$ channels even in future long baseline experiments. However, if one is able to reach absolute precision of $\sim 1\%$ in probability, then it is important to consider these terms alongside the effects of $\gamma_3$.
 	
 	\item The diagonal components of the general decay matrix $\gamma_1$ and $\gamma_2$ are absent in $P_{e\mu}^{(\Gamma)}$, whereas the contribution from the off-diagonal elements $\gamma_{13}$ and $\gamma_{23}$ are present.
 \end{itemize}

Note that the vacuum limits with the inclusion of general decay matrix $\Gamma$ match with our results obtained via the 3-flavor Zassenhaus expansion method in section~\ref{sec:zassenhaus1}.

\section{Comparison of analytical expressions with numerical results}
\label{sec:numcom}
In this section, we compare the accuracy of our analytic approximations against the exact numerical results. For this, the 3-neutrino mixing parameter values taken are
\begin{align}
	\theta_{12}=33^\circ \;,\qquad	\theta_{23}\simeq 45^\circ\;,\qquad &	\theta_{13}\simeq 8.5^\circ\;,\qquad \delta_{\text{CP}} =0^\circ\;, \qquad \nonumber\\
	\Delta m_{21}^2 =7.37\times10^{-5}\; \text{eV}^2\;, \qquad& \Delta m_{31}^2 = 2.56\times 10^{-3}\;\text{eV}^2\;.
\end{align}
These values are consistent with the global fit \cite{nufit,Esteban:2020cvm,deSalas:2020pgw,Capozzi:2021fjo} within $3\sigma$ for normal mass ordering ($\Delta m^2_{31}>0$) of neutrinos.

We present our results for $P_{\mu\mu}$ and $P_{\mu e}$, the most relevant probabilities for long baseline experiments. For the purpose of illustration, we take the simpler case of only $\nu_3$ decaying, by choosing $\gamma_3=0.1$ and all the other elements of the decay matrix $\Gamma$ equal to be zero in vacuum. Note that the presence of matter will give rise to off-diagonal elements of the decay matrix $\Gamma_m$ in the matter basis, which is automatically taken care of in our analysis.

For the  scenario without decay, we show the comparison of the numerical results with two analytic approximations:
\begin{itemize}
	\item \textit{OMSD}: $P_{\mu e}$ and $P_{\mu\mu}$ calculated in the OMSD approximation (with exact dependence on $s_{13}$).
	\item \textit{Expansion}: $P_{\mu e}$ calculated up to $O(\lambda^3)$ and $P_{\mu\mu}$ calculated up to $O(\lambda^2)$ in $s_{13}$ and  $\alpha$.
\end{itemize} 

For the scenario with decay, we show the comparison of the numerical results with three analytic expansions:
\begin{itemize}
	\item \textit{OMSD}: $P_{\mu e}$ and $P_{\mu\mu}$ calculated in the OMSD approximation, with exact dependence on $s_{13}$, $\gamma_1^m$, $\gamma_3^m$ and $\gamma_{13}^m$ (Note that the non-zero $\gamma_3$ leads to non-zero $\gamma_1^m$, $\gamma_3^m$ and $\gamma_{13}^m$ in the OMSD basis).
	\item \textit{Expansion}: $P_{\mu e}$ calculated up to $O(\lambda^3)$ and $P_{\mu\mu}$ calculated up to $O(\lambda^2)$ in $s_{13}$ and $\alpha$, with exact dependence on $\gamma_3$.
	\item \textit{Full-Expansion}: $P_{\mu e}$ calculated up to $O(\lambda^3)$ and $P_{\mu\mu}$ calculated up to $O(\lambda^2)$ in $s_{13}$, $\alpha$ and $\gamma_3$.

\end{itemize}
Note that Expansion is always an improvement analytically over Full-Expansion. We quantify the accuracy of these analytic approximations in terms of the quantity
\begin{equation}
	\Delta P_{\alpha \beta} = P_{\alpha \beta}(\text{analytic} )- P_{\alpha \beta}(\text{numerical} )\;.
\end{equation}

We first show the comparison for the next-generation long-baseline neutrino experiment DUNE with $L=1300$ km, and then for a hypothetical ``magic-baseline''~\cite{Huber:2003ak} experiment with $L=7000$ km.
Later we discuss the utility of our analytic approximations over a wide range of energies and baselines.

\subsection{At a baseline $L=1300$ km}
\label{sec:1300}
The future long-baseline experiment DUNE has the energy range $E_\nu \simeq 0.5 - 10$ GeV, and baseline $L \simeq1300$ km. The oscillation probability, when the detailed earth density profile is taken into account, may be reproduced very accurately when the approximation of a constant density $\rho_{\text{avg}} \simeq 3$ g/cc along the path of the neutrino is used.
Our numerical results have been calculated with this constant density approximation.

\begin{figure}[t]\centering
	\hspace{5pt} \includegraphics[width=0.465\textwidth]{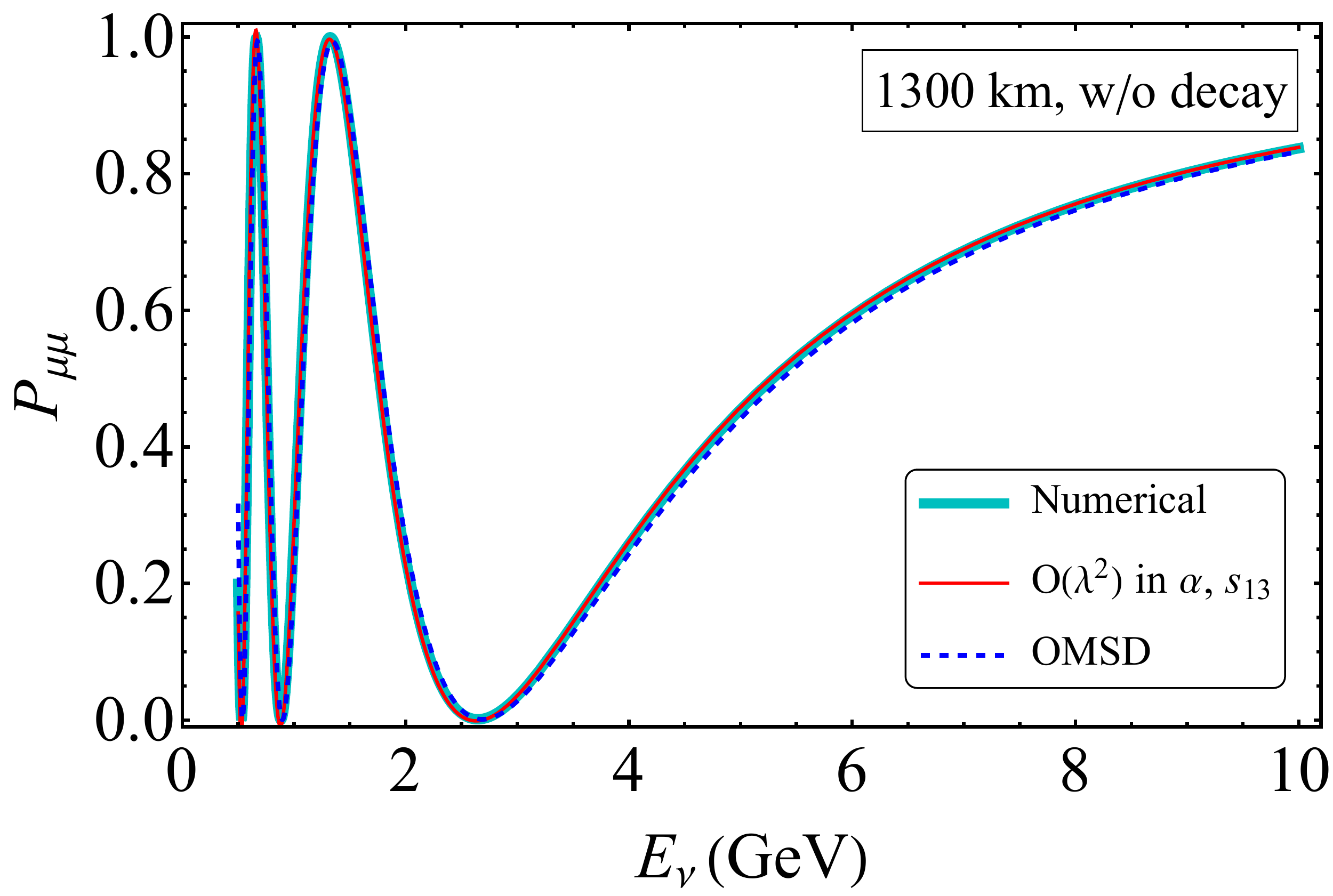} \hspace{3.2pt}
	\includegraphics[width=0.465\textwidth]{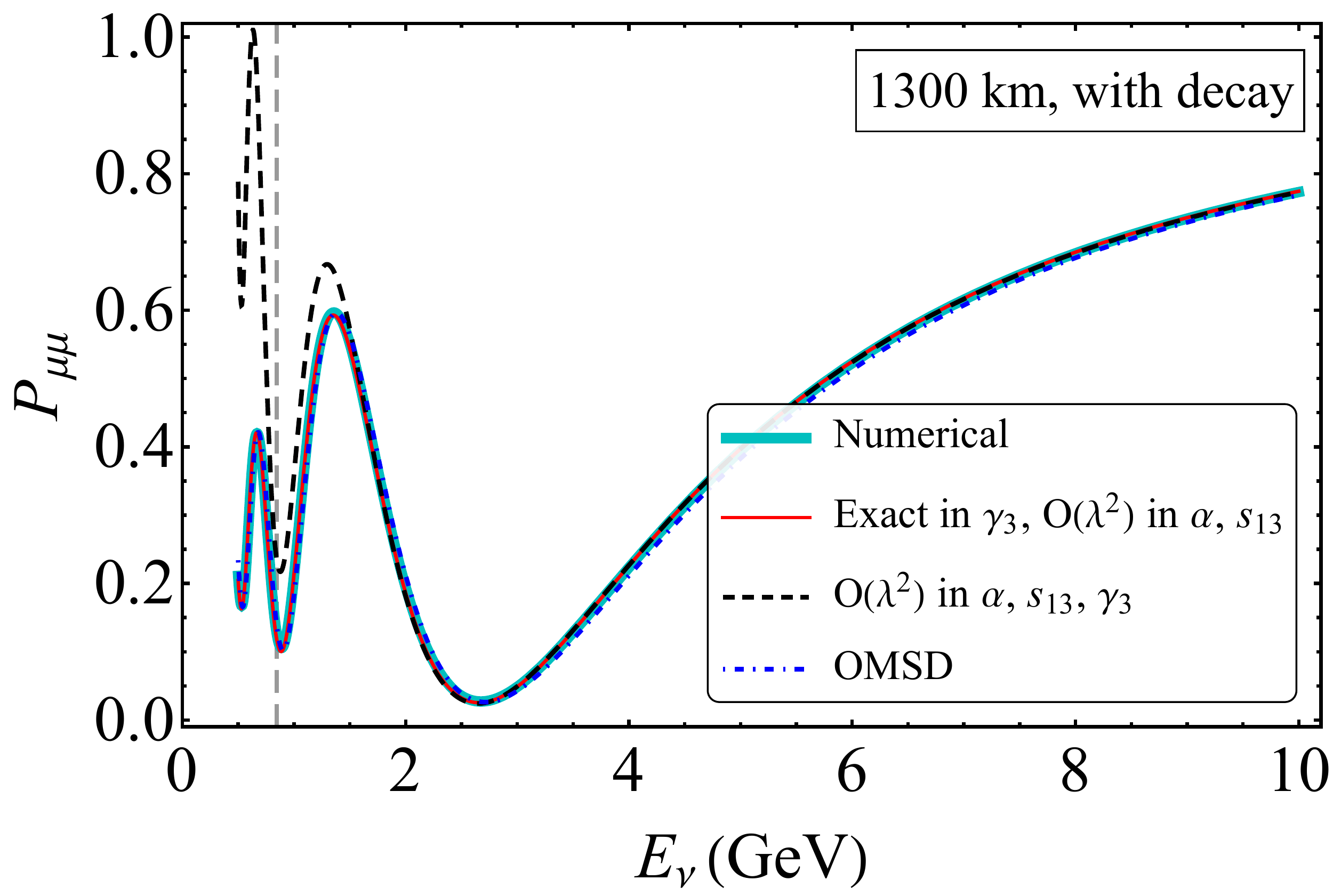} \hspace{7pt} \\
	\vspace{10pt}
	\includegraphics[width=0.48\textwidth]{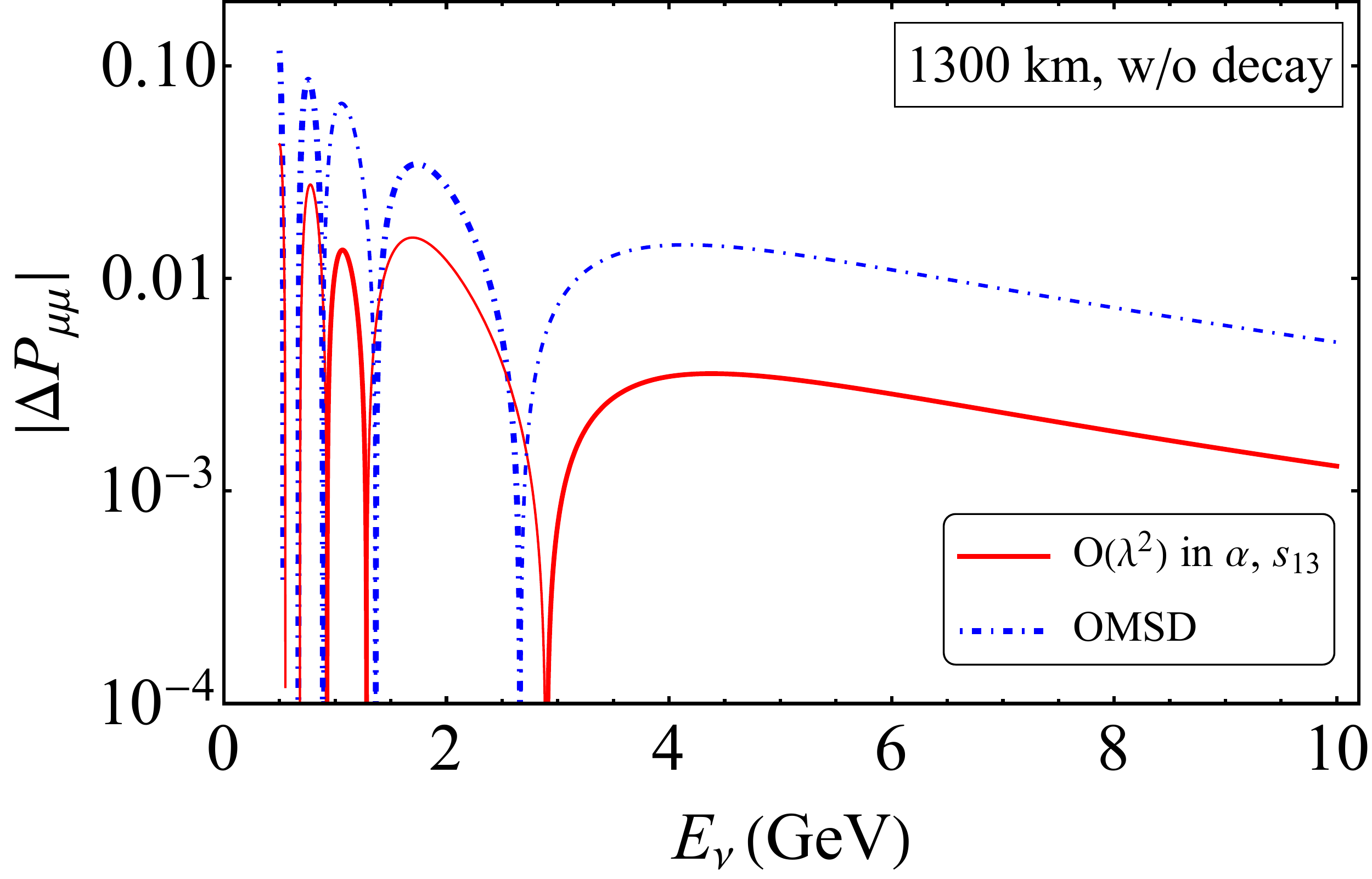}
	\includegraphics[width=0.48\textwidth]{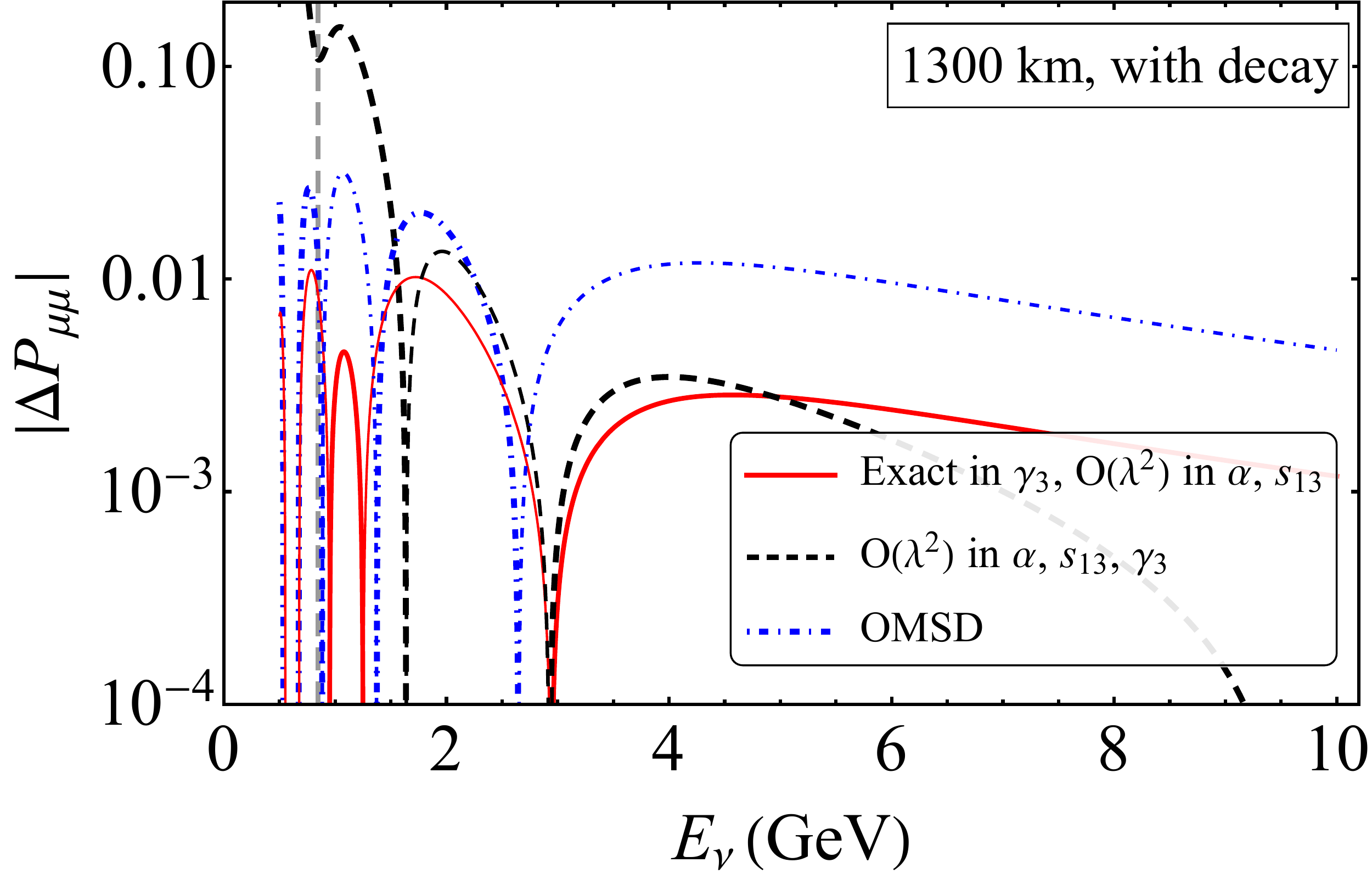}
	\caption{The top panels show probabilities $P_{\mu \mu}$ in the scenarios without (left) and with (right) decay, for $L=1300$ km and  $\gamma_3=0.1$. The bottom panels show the  absolute error $|\Delta P_{\mu\mu}|$ for the analytic expressions shown. The thick (thin) curves indicate positive (negative) signs of $\Delta P_{\mu\mu}$. The dashed vertical line at $E_\nu \simeq 0.8$ GeV corresponds to $\lambda \Delta =1$, to the left of which the expansion in $\gamma_3$ is not expected to be valid.
	} 
	\label{fig:1}
\end{figure}
In figure~\ref{fig:1}, we compare the probabilities $P_{\mu \mu}$ obtained from the analytic expressions with the exact numerical results, in the scenarios ``without decay'' and ``with decay''. We can make the following observations:
\begin{itemize}
	\item For the scenario without decay, analytic expressions --- both with OMSD as well as the Expansion --- reproduce the dip and peak positions quite accurately. 
	The probability of the first oscillation peak as well as the dip (the ones at highest energies) is also very well reproduced.
	
	\item For the scenario without decay, the Expansion method gives the probability with $|\Delta P_{\mu\mu }| \lesssim 1\%$ for energies $E_\nu >1$ GeV. This is expected from results in earlier literature. 
	
	\item For the scenario with decay, the height of the first oscillation peak reduces all the way from 1 to 0.6 (for $\gamma_3=0.1$). Our analytic expressions predict this, since the contribution of $\gamma_3$ to $P_{\mu\mu}$ has been shown to be $O(\lambda)$. This could, therefore, be one of the most prominent signatures of the decay of $\nu_3$.
	
	\item The positions of dips and peaks are predicted quite accurately by all our approximations. The Expansion method (the dependence on $\gamma_3$ exactly calculated) gives an accuracy of $|\Delta P_{\mu\mu }| \lesssim 1\%$ for the whole energy range of 0.5 -- 10 GeV. 
	
	\item Note that for the first as well as second oscillation dip (as counted from the highest energies) the probability $P_{\mu\mu}$ in the scenario with decay gets a non-zero value, though it was zero for the scenario without decay. This is a curious observation, although one that can be explained via our analytic expressions. The explanation as well as the implications of this will be further expanded upon on section~\ref{sec:oscdip}.
\end{itemize}

\begin{figure}[t]\centering
	\includegraphics[width=0.48\textwidth]{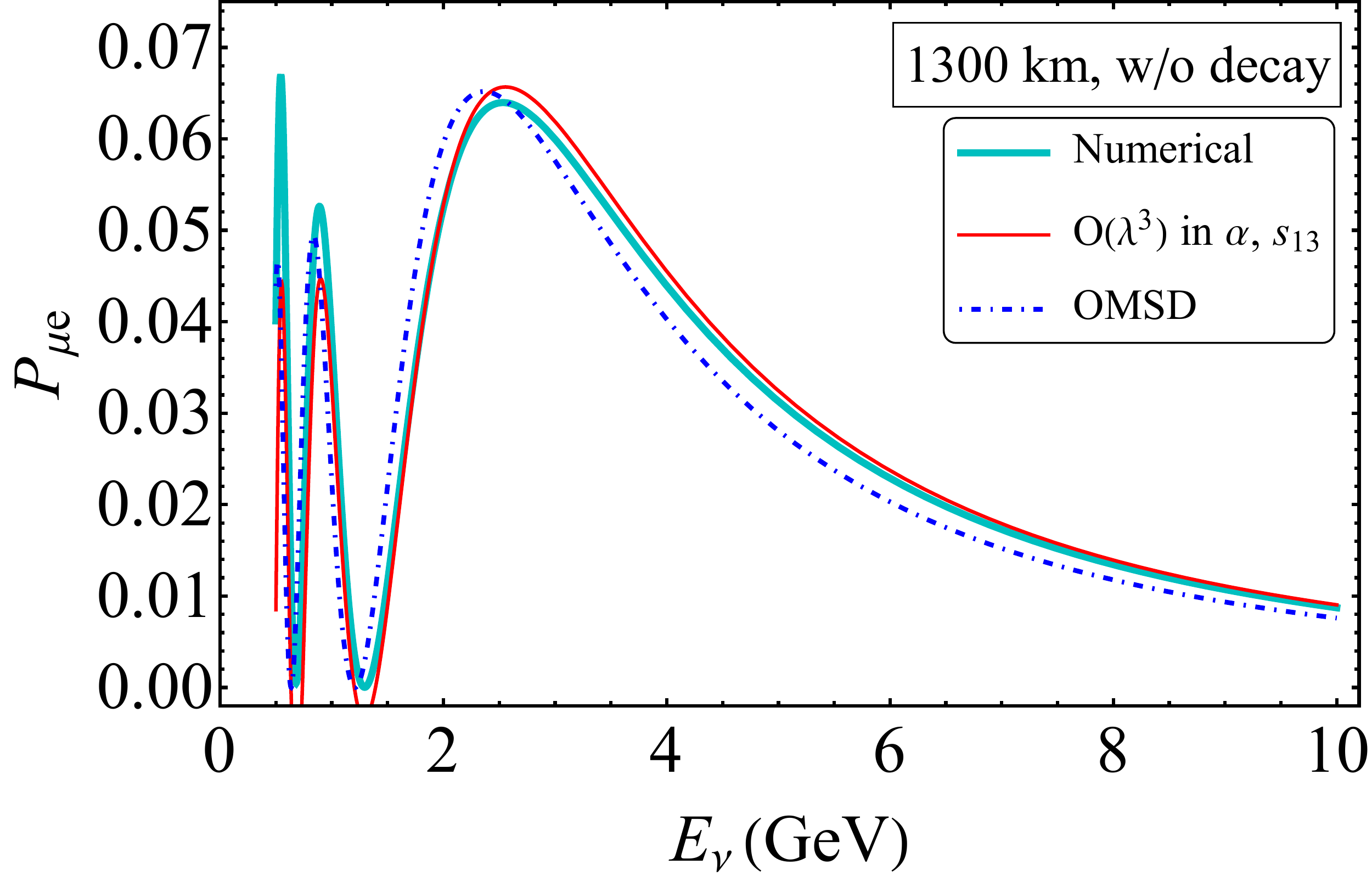}
	\includegraphics[width=0.48\textwidth]{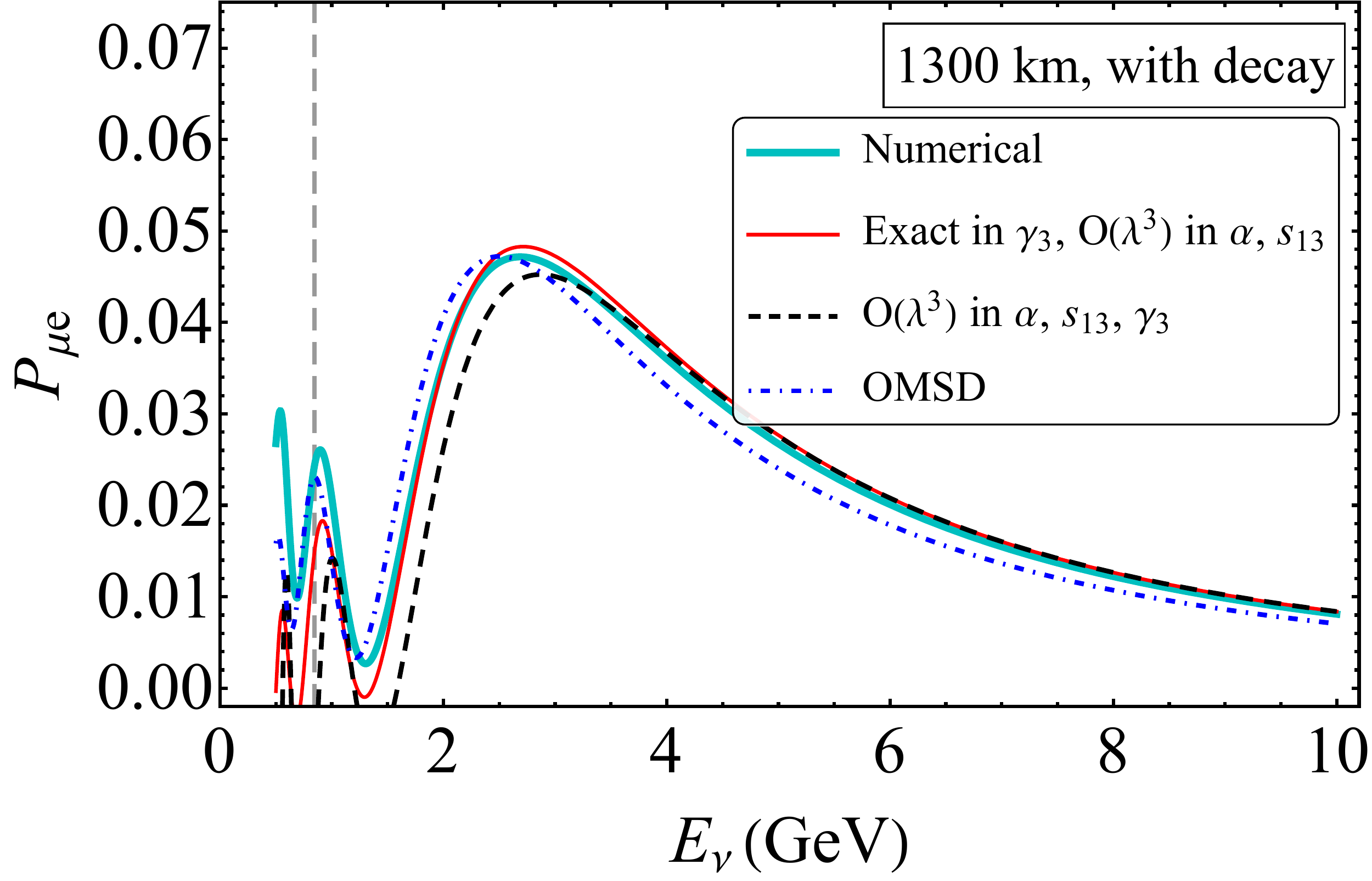}\\
	\vspace{10pt}
	\includegraphics[width=0.48\textwidth]{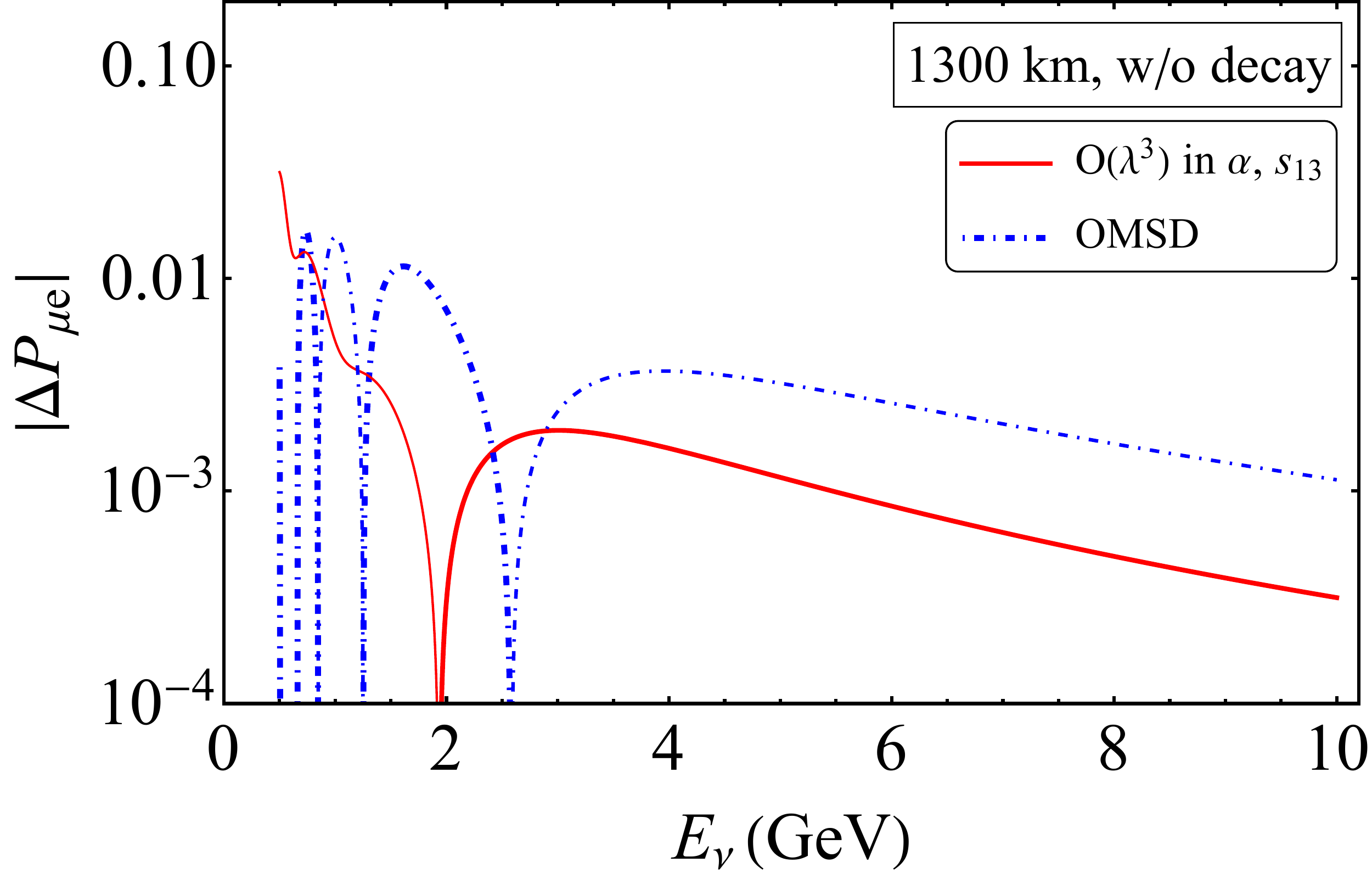}
	\includegraphics[width=0.48\textwidth]{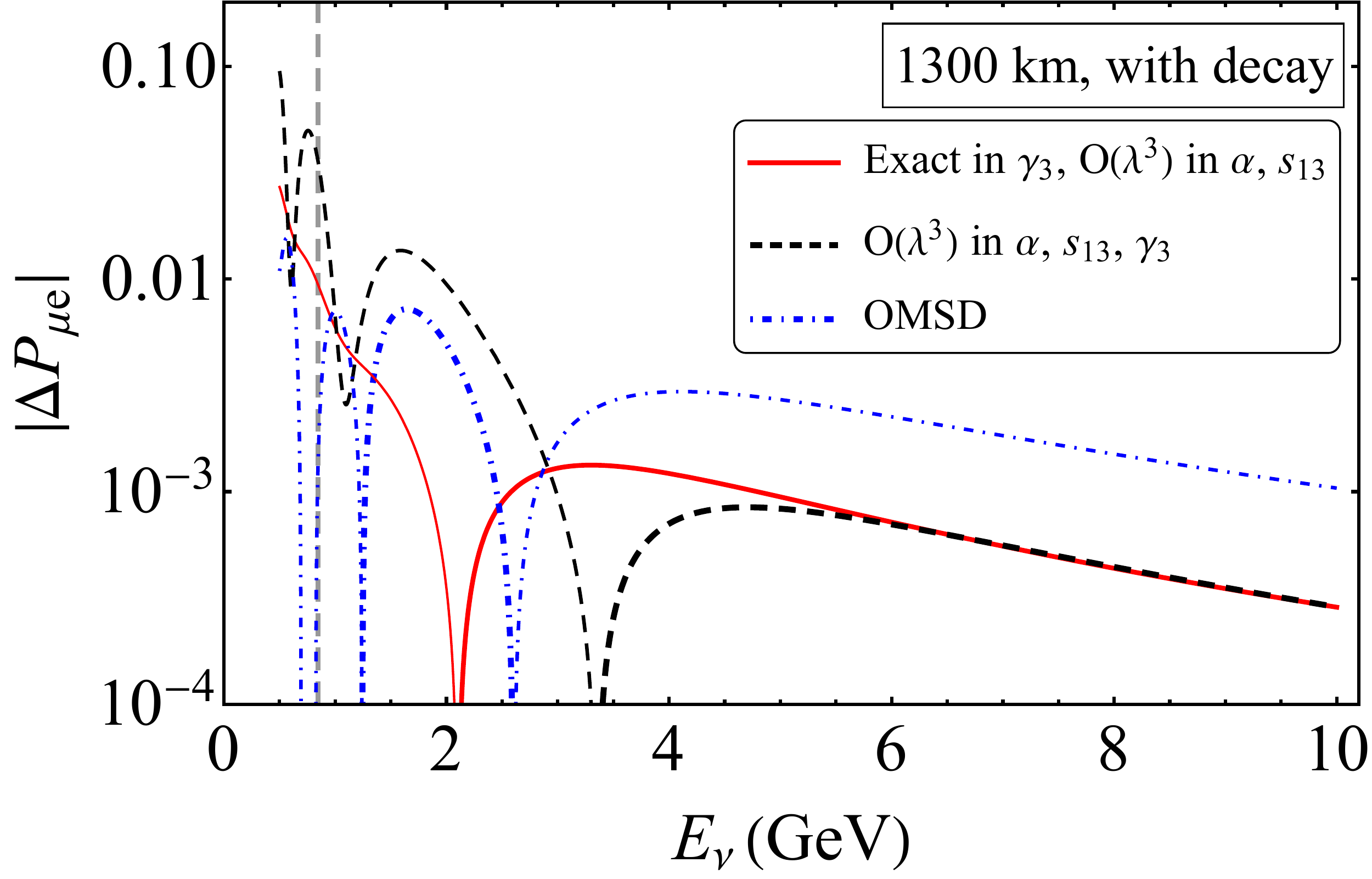}
	\caption{
		The top panels show probabilities $P_{\mu e}$ in the scenarios without (left) and with (right) decay, for $L=1300$ km and  $\gamma_3=0.1$. The bottom panels show the  absolute error $|\Delta P_{\mu e}|$ for the analytic expressions shown. The thick (thin) curves indicate positive (negative) signs of $\Delta P_{\mu e}$. The dashed gray vertical line follows the same convention as that mentioned in figure~\ref{fig:1}.} 
		\label{fig:2}
\end{figure}

Our analytic approximations in the both the scenarios with and without decay become less accurate become less accurate for lower values of energies (i.e. higher values of $\Delta$). This is expected since at lower energies we approach the regime where $\Delta m_{21}^2$ oscillation is also important or the term describing decay is no longer linear in $\gamma_3 \Delta$ (the latter only affects the Full-Expansion results).

In figure~\ref{fig:1} (and also in later figures), we observe many sharp dips in the $|\Delta P_{\alpha\beta}|$ at many discrete values of energies indicating that $|\Delta P_{\alpha\beta}| =0$. This should not be taken as a measure of absolute accuracy since this simply happens at the energies where the values from analytic expressions coincidentally match the numerical results (the curves corresponding to analytic expressions ``cross'' the curves obtained numerically).  

In figure~\ref{fig:2}, we compare the probabilities $P_{\mu e}$ obtained from the analytic expressions with the exact numerical results, in the scenarios without decay and with decay. We can make the following observations:

\begin{itemize}
	\item As in the case of $P_{\mu\mu}$, the dip and peak positions in both the scenarios without and with decay are reproduced quite accurately in all approximations.
	
	\item The values of probabilities are reproduced very well  with OMSD as well as the Expansion method for the scenarios without and with decay, with $|\Delta P_{\mu e }| \lesssim 1\%$ for energies $E_\nu >1$ GeV.
	
	\item For the scenario with decay, the height of the first oscillation peak reduces from $\approx 0.065$ to $\approx 0.050$ (for $\gamma_3=0.1$). Since the absolute decrease in probability is quite small, it would be difficult to isolate the effect of decay in $P_{\mu e}$ channel for this baseline. This is expected since our analytic expression show that the leading term in $P_{\mu e}$ is itself $O(\lambda^2)$, while the leading order modification due to $\gamma_3$ is at $O(\lambda^3)$.
\end{itemize}

Since the accuracy of our analytic expressions is better than $1\%$ in probability and the accuracy of probability measurement at an experiment like DUNE may reach $\sim 1\%$, one may be able to identify the effect of neutrino decay, which may lead to $O(\lambda)$ modifications in the probability $P_{\mu\mu}$.

\subsection{At a baseline $L=7000$ km}
\label{sec:7000}
The magic baseline of $L=7000$ km  \cite{Barger:2001yr,Huber:2003ak} has been discussed extensively in the literature, since the probabilities at this baseline are almost independent of the unknown CP phase $\delta_{\text{CP}}$. As a result, the 3-neutrino oscillation problem may be crafted into an effective 2-neutrino oscillation problem using the OMSD approximation, which is detailed in section~\ref{sec:OMSD}. The OMSD approximation is expected to work very well in this regime, as has been pointed out earlier \cite{Gandhi:2004bj}. Although no experiment is currently planned with this magic baseline, we present our results at this baseline to point out the power of OMSD approximation even when neutrinos decay.

We take the energy range of $E_\nu \simeq 2-25$ GeV. At $L=7000$ km, the constant average density approximation of the Earth is expected to yield quite accurate neutrino probabilities with the PREM-averaged density $\rho_{\text{avg}}=4.15$ g/cc  \cite{Gandhi:2004bj}. 
In figure~\ref{fig:3} we compare the probabilities $P_{\mu \mu}$ obtained from the analytic expressions with the exact numerical results in the average density approximation, in the scenarios without and with decay, for $L=7000$ km. The following observations can be made:

\begin{figure}[t]\centering
	\hspace{0.45pt} \includegraphics[width=0.466\textwidth]{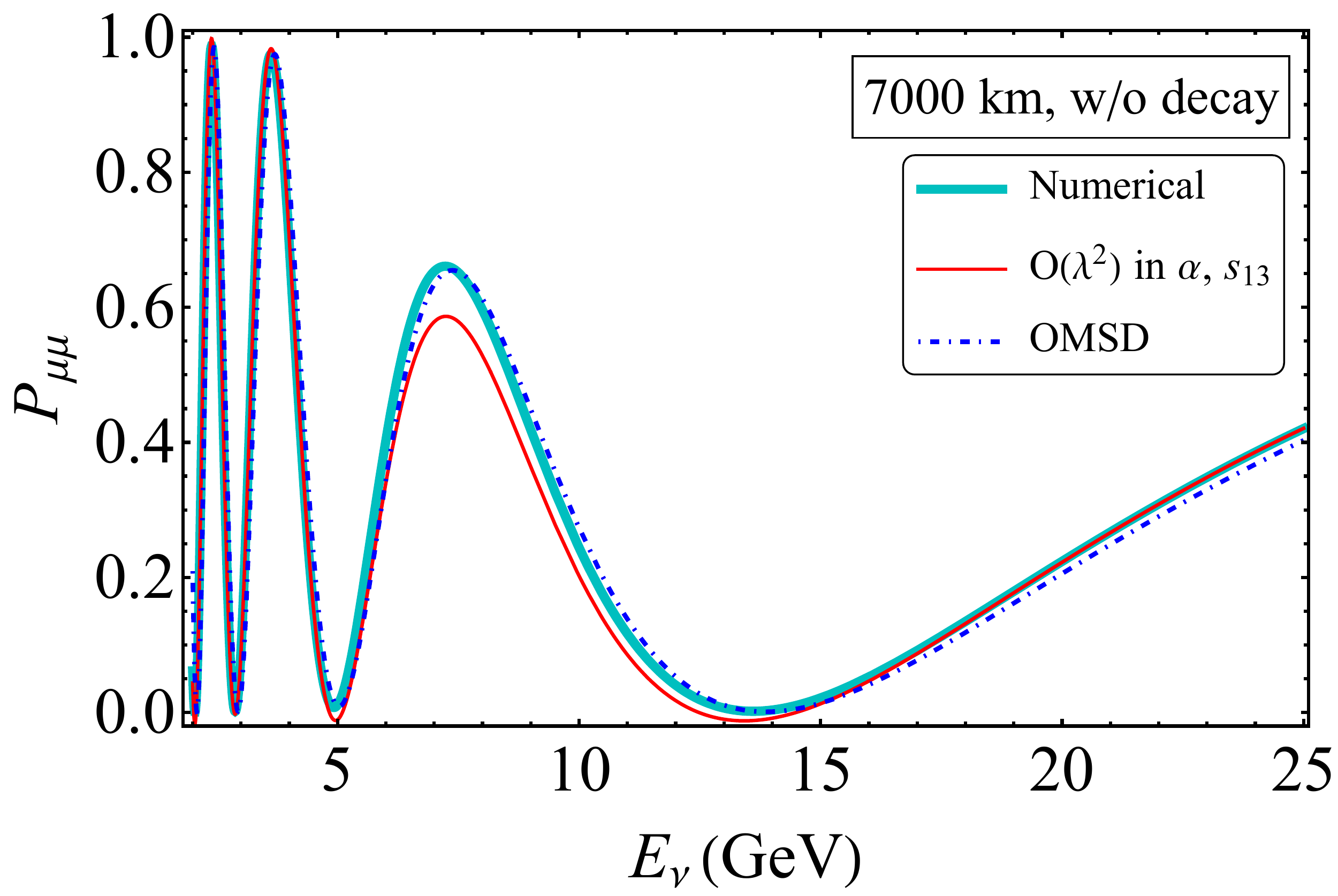} \hspace{2.9pt}
	\includegraphics[width=0.466\textwidth]{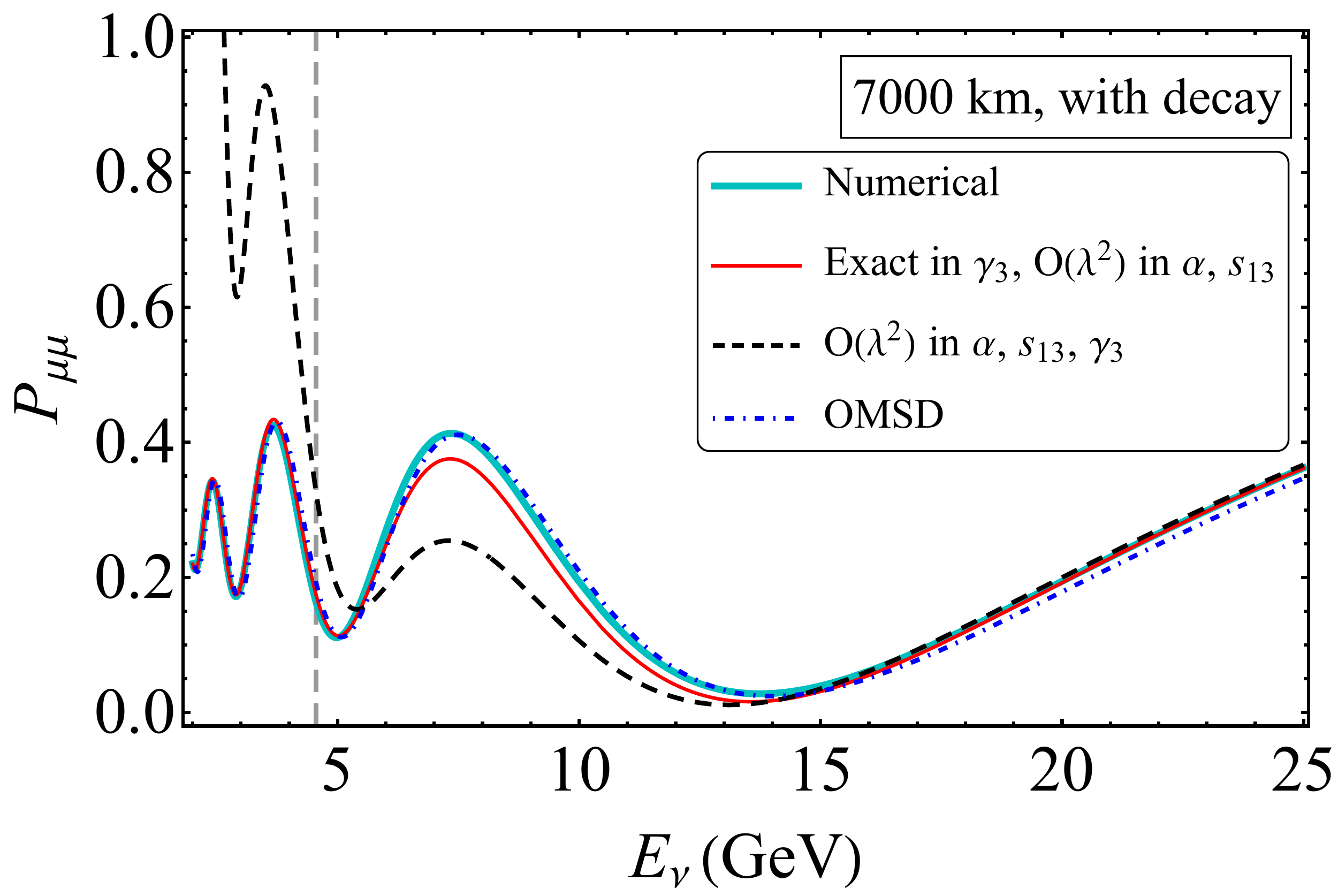} \hspace{7pt}\\
	\vspace{10pt}
	\includegraphics[width=0.48\textwidth]{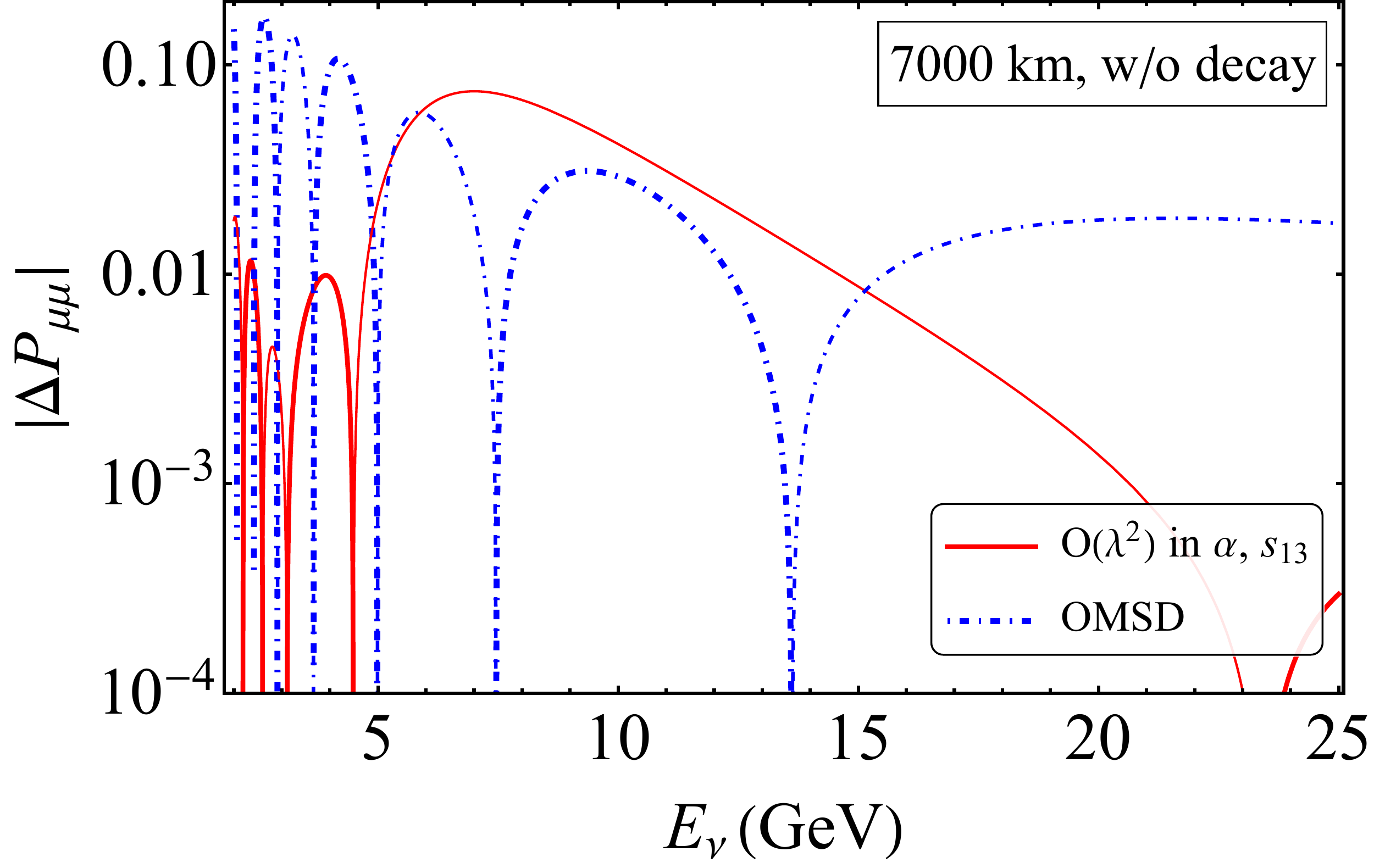}
	\includegraphics[width=0.48\textwidth]{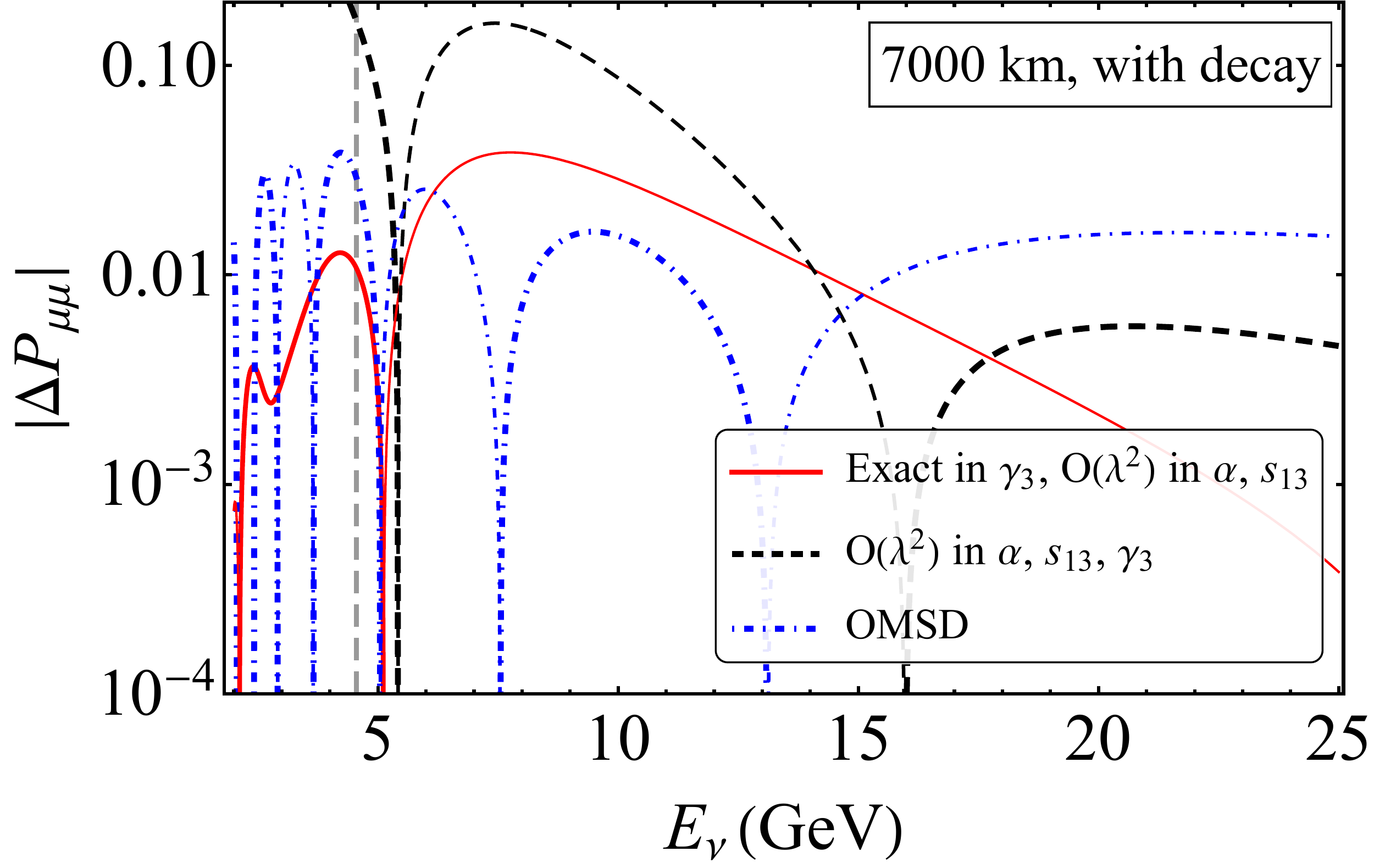}
	\caption{
		The top panels show probabilities $P_{\mu \mu}$ in the scenarios without (left) and with (right) decay, for $L=7000$ km and  $\gamma_3=0.1$. The bottom panels show the  absolute error $|\Delta P_{\mu\mu}|$ for the analytic expressions shown. The thick (thin) curves indicate positive (negative) signs of $\Delta P_{\mu\mu}$. The dashed vertical line at $E_\nu \simeq 4.5$ GeV corresponds to $\lambda \Delta =1$, to the left of which the expansion in $\gamma_3$ is not expected to be valid.} 
	\label{fig:3}
\end{figure}
\begin{itemize}
	\item For the scenario without decay, analytic expressions --- both with OMSD as well as the Expansion --- reproduce the position of dips and peaks of the oscillation quite well. The OMSD approximation especially reproduces the heights of the first oscillation peak and the dip (from high to low energy) very accurately. This is expected behavior as seen in earlier literature.
	
	\item For the scenario without decay, the Expansion method gives the probability with $|\Delta P_{\mu\mu }| \lesssim 4\%$ in all energies of interest except for $E_\nu \simeq 5 - 10$ GeV. The OMSD approximation yields results accurate to $|\Delta P_{\mu\mu }| \lesssim 4 \%$  for all $E_\nu \gtrsim 7$ GeV.
	
	\item The oscillation peak (counting from high to low energies) may be seen to have shifted marginally. This gives rise to the seemingly large values of $|\Delta P_{\mu\mu}|$. However, given the expected errors on the measurement of neutrino energy this shift of 0.1 -- 0.2 GeV is quite negligible.
	
	\item For large energies the OMSD approximation is always below the numerically calculated probability by $\Delta P_{\mu\mu } \approx 2 \%$ in the energy range shown. This is primarily because the OMSD approximation ignores the contribution due to $\Delta m_{21}^2$. The error in the Expansion method however goes down to $\ll 1\%$ at higher energies.
	
	\item For the scenario with decay, the height of the first oscillation peak reduces from 0.65 to 0.4 (for $\gamma_3=0.1$). This is in line with the expected modifications at the leading order that arise due to $\nu_3$ decay. We also observe significant reduction of probabilities at the second and third oscillation peaks from $\approx 0.95$ to $\approx 0.4$, which may prove to be a promising signature of neutrino decay.
	
	\item The positions of dips and peaks are predicted quite accurately by all our approximations. The OMSD as well as the Expansion method (the dependence on $\gamma_3$ exactly calculated) gives an accuracy of $|\Delta P_{\mu\mu }| \lesssim 4\%$ for the whole energy range.
	
	\item Again, similar to the observations made for $P_{\mu\mu }$ in the previous subsection, for the first as well as second oscillation dip (as counted from the highest energies) the probabilities in the scenario with decay become non-zero. This interesting feature is observed to be more prominent at $7000$ km baseline, with as much as $10\%$ of muon neutrinos surviving at the first oscillation dip.
\end{itemize}
\begin{figure}[t]\centering
	\hspace{4.pt}\includegraphics[width=0.4665\textwidth]{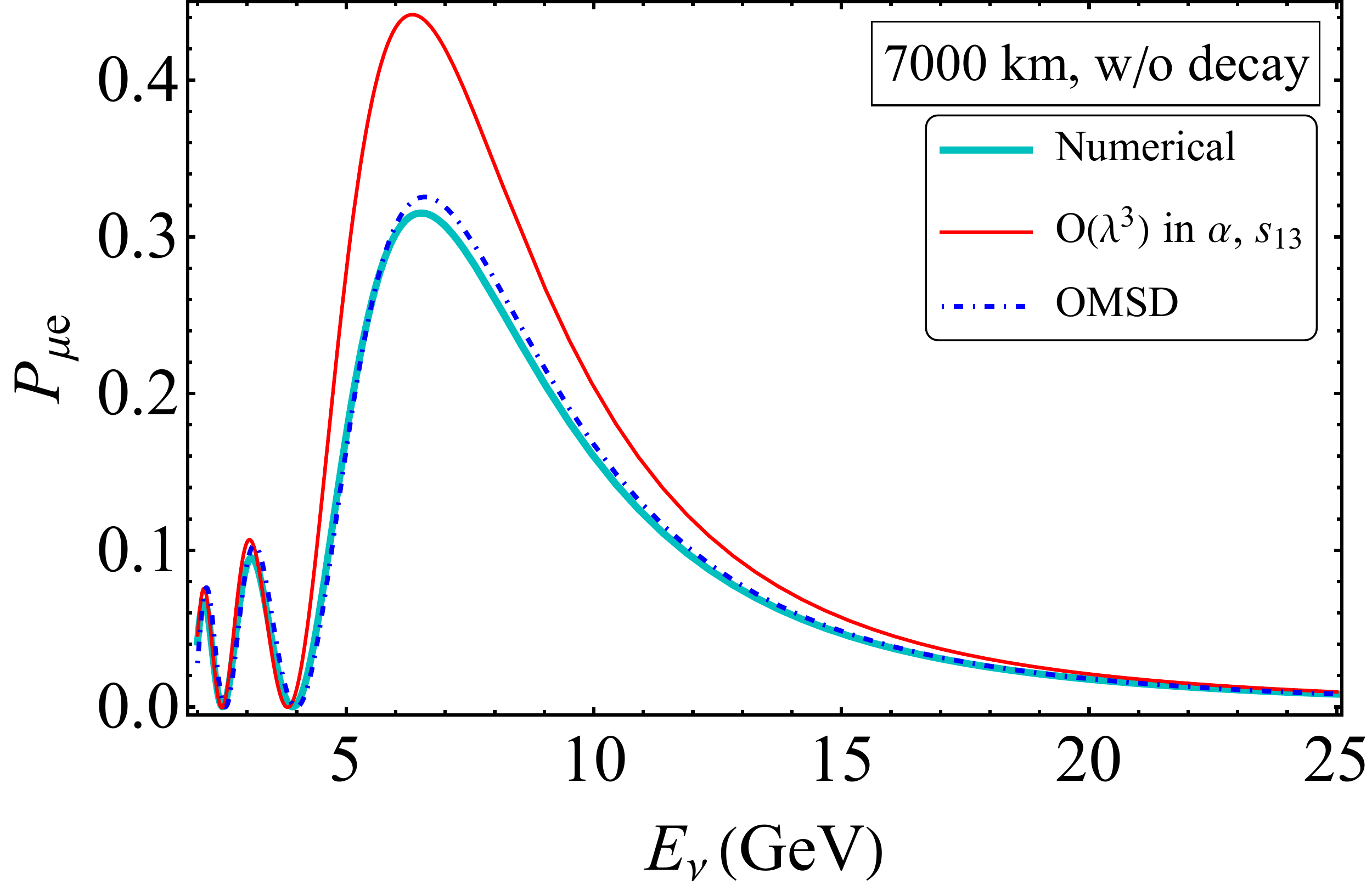} \hspace{2.6pt}
	\includegraphics[width=0.465\textwidth]{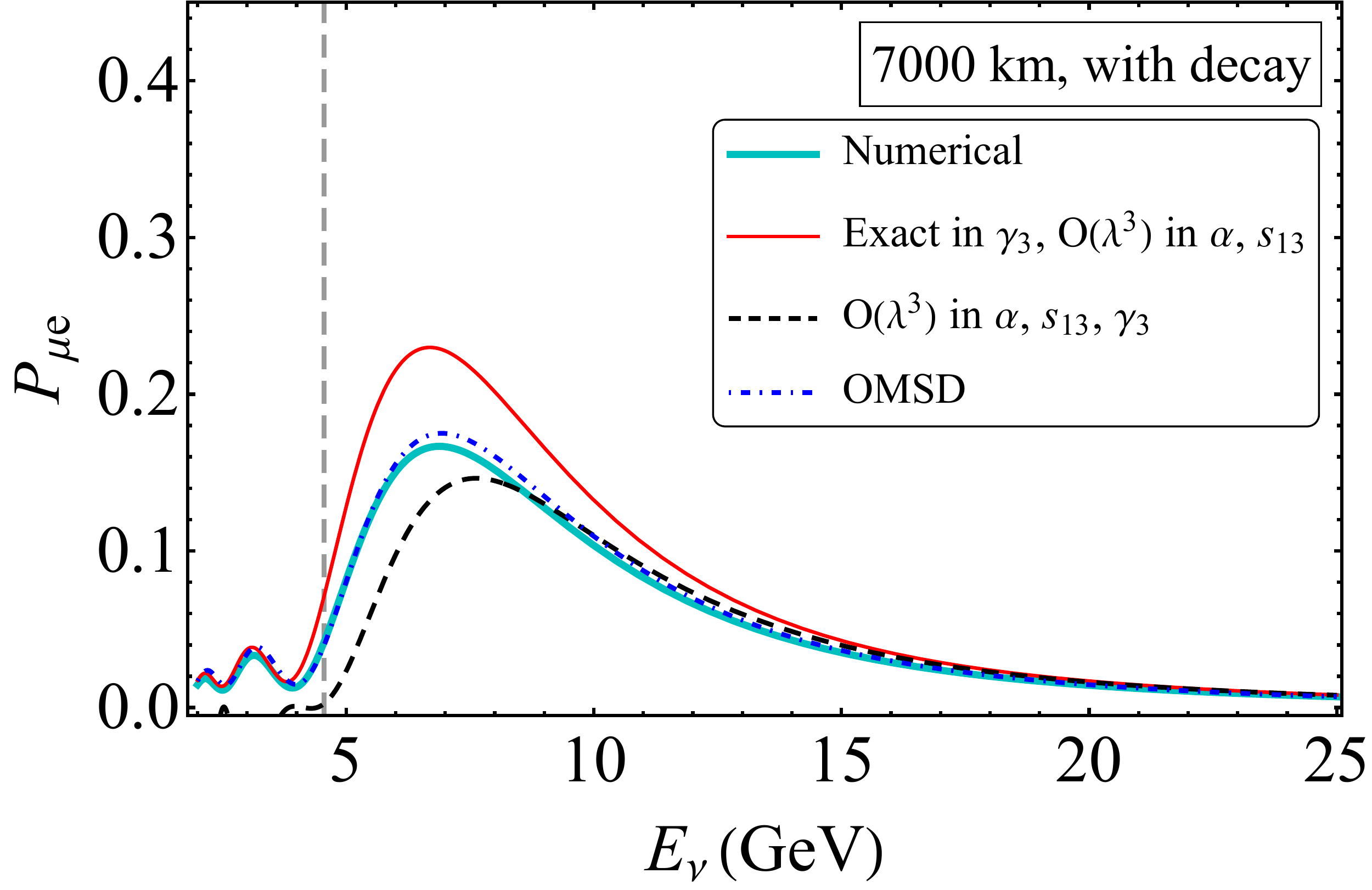}\hspace{2pt}\\
	\vspace{10pt}
	\includegraphics[width=0.48\textwidth]{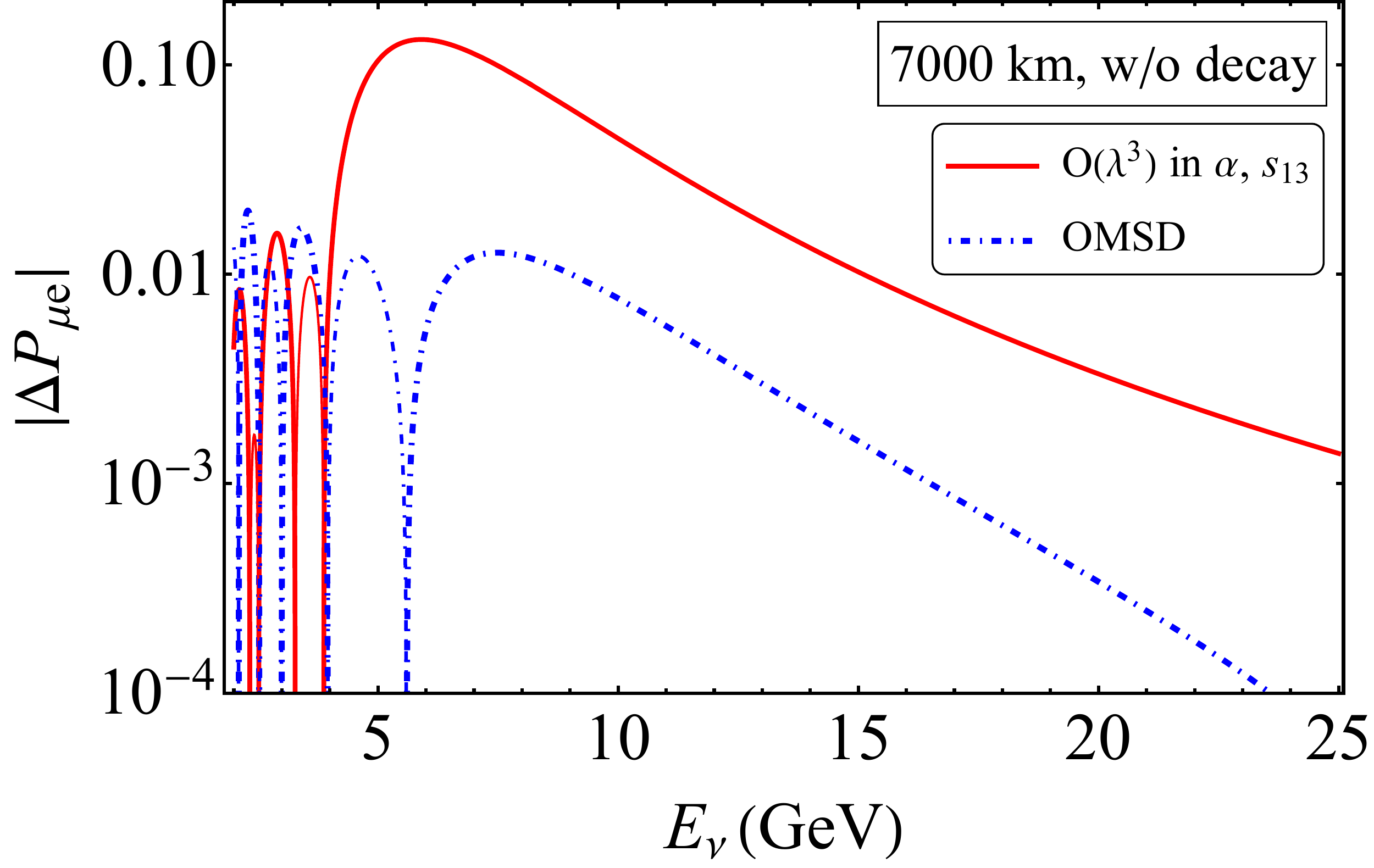}
	\includegraphics[width=0.48\textwidth]{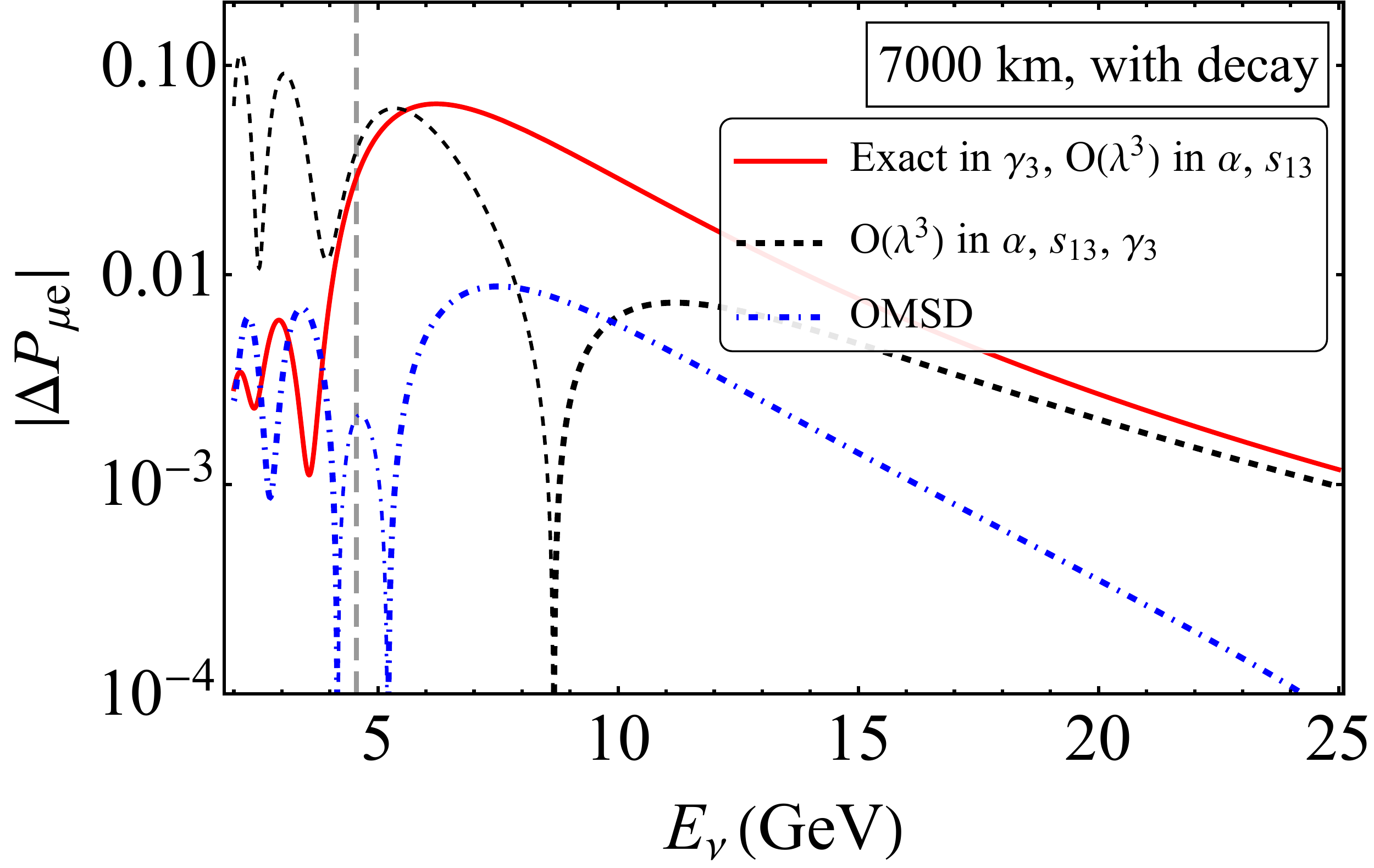}
	\caption{
		The top panels show probabilities $P_{\mu e}$ in the scenarios without (left) and with (right) decay, for $L=7000$ km and  $\gamma_3=0.1$. The bottom panels show the  absolute error $|\Delta P_{\mu e}|$ for the analytic expressions shown. The thick (thin) curves indicate positive (negative) signs of $\Delta P_{\mu\mu}$. The dashed gray vertical line follows the same convention as that mentioned in figure~\ref{fig:3}.} 
	\label{fig:4}
\end{figure}

In figure~\ref{fig:4}, we compare the probabilities $P_{\mu e}$ obtained from the analytic expressions with the exact numerical results, in the scenarios without and with decay. We can make the following observations:
\begin{itemize}
	\item Similar to all the previous cases, the dip and peak positions in both the scenarios without and with decay are reproduced quite well in all approximations. The OMSD approximation also predicts the heights of the oscillation peaks and dips quite well.

	\item The probabilities are reproduced very well with the OMSD approximation; $|\Delta P_{\mu\mu }| \lesssim 1\%$ for all energies, with or without decay. 
	
	\item The Expansion method predicts too large a conversion probability and hence is not suitable for comparing with precision measurements of $P_{\mu e}$ at $L=7000$ km. 
	
	\item For the scenario with decay, the height of the first oscillation peak reduces from $\approx 0.3$ to $\approx 0.16$ (for $\gamma_3=0.1$). This is a quite significant decrease and definitely within the domain of measurability if indeed $\gamma_3 \sim O(0.1)$.
\end{itemize}
To summarize, at the magic baseline of $L=7000$ km, both $P_{\mu\mu}$ and $P_{\mu e}$ would serve as channels to look for neutrino decay and the OMSD approximation works extremely well at the magic baseline for explaining the features of the probabilities.

\subsection{Over a wide range of baselines}

\begin{figure}[t]\centering
	\includegraphics[width=0.48\textwidth]{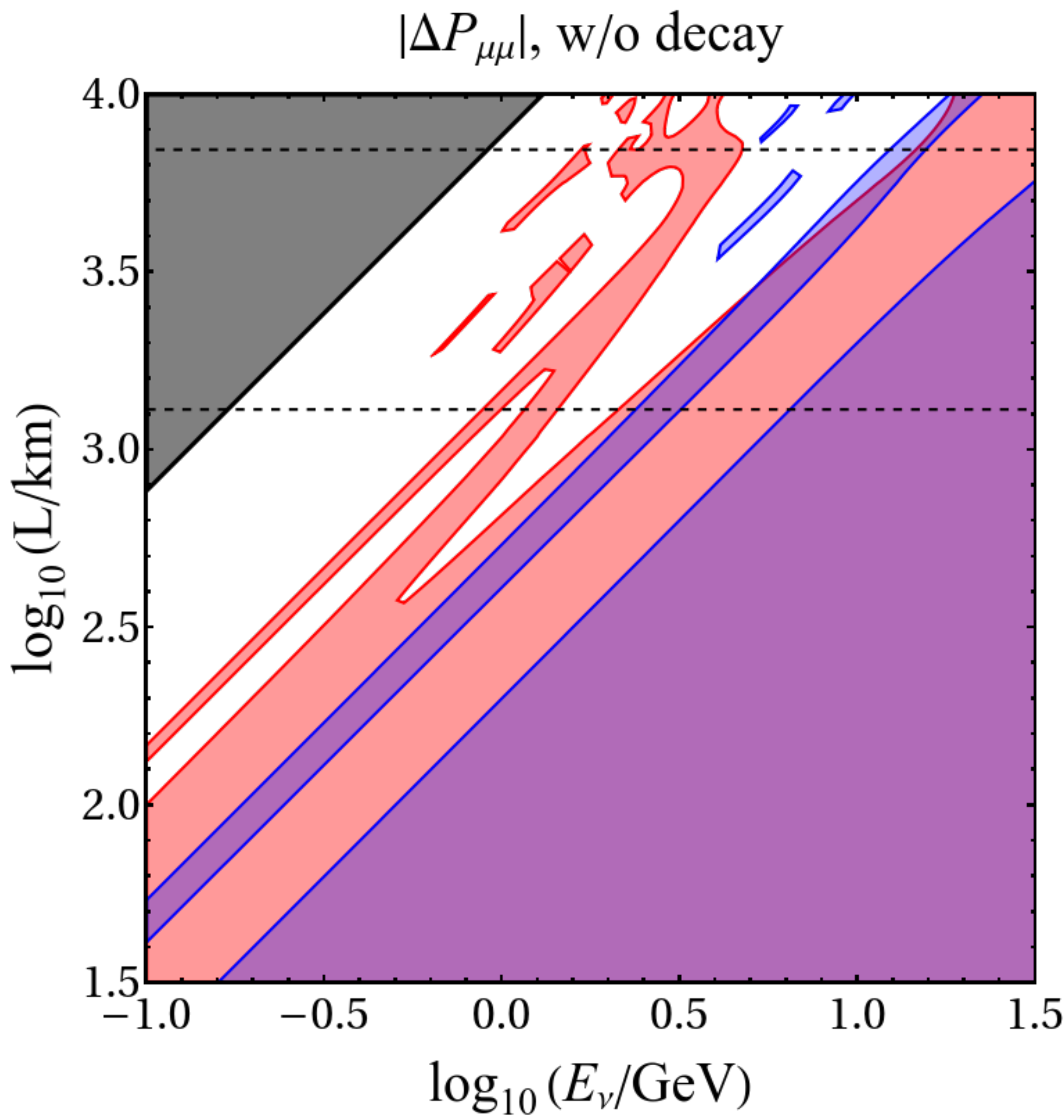}
	\includegraphics[width=0.48\textwidth]{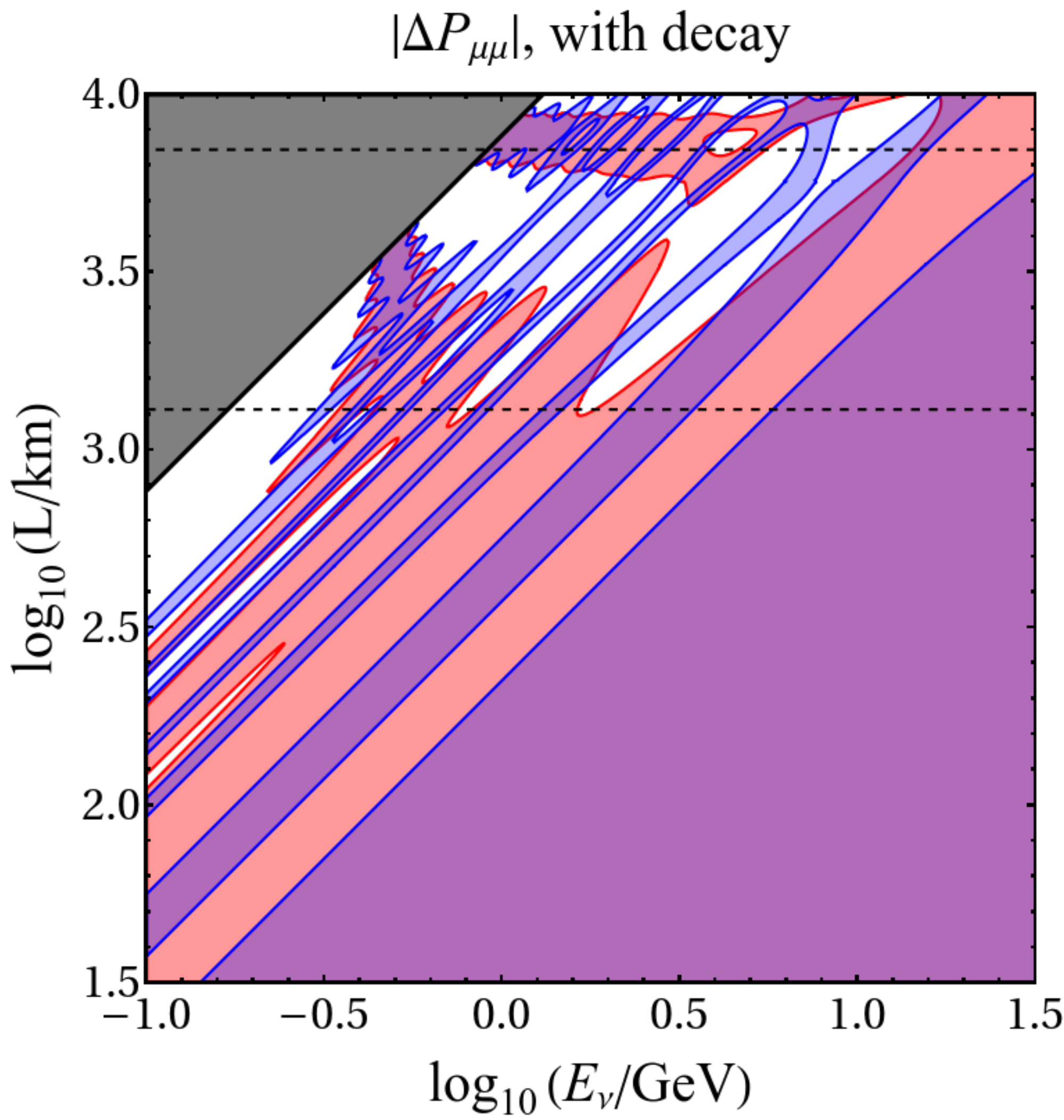}\\
	\vspace{10pt}
	\includegraphics[width=0.48\textwidth]{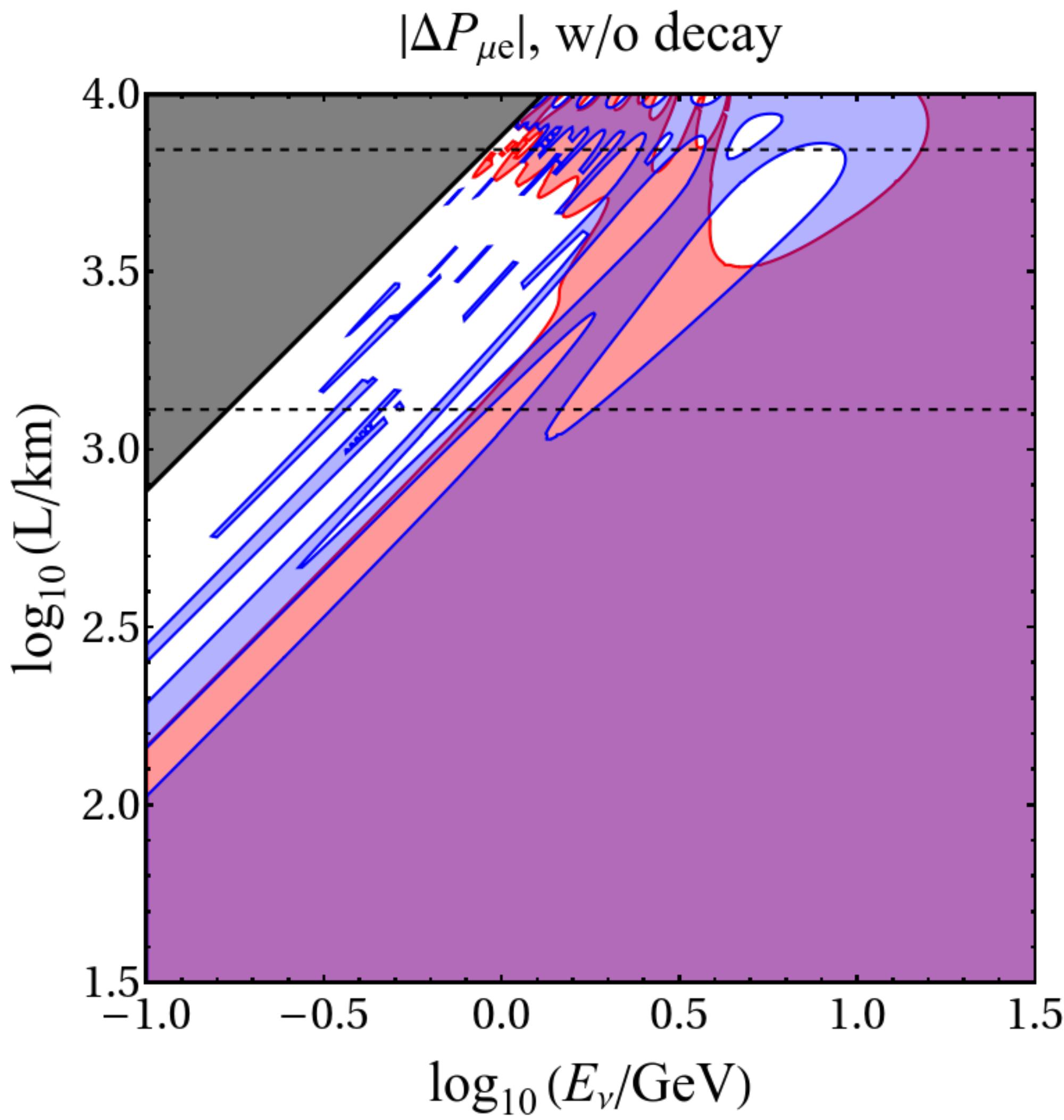}
	\includegraphics[width=0.48\textwidth]{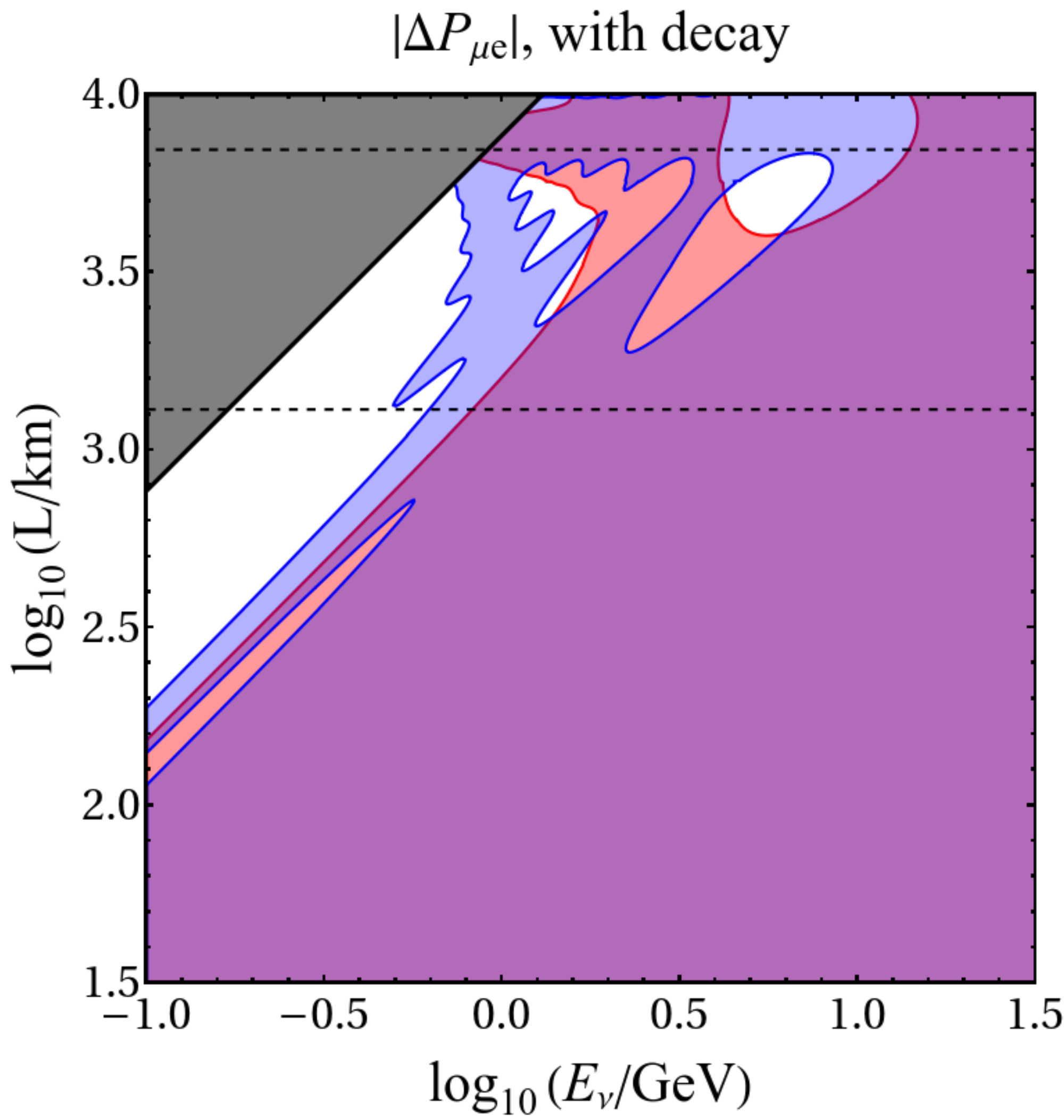}
	\caption{
		The regions in the $(E_\nu,L)$ parameter space where $|\Delta P_{\alpha\beta}| <1\%$ with the OMSD approximation (blue) and the Expansion method (red).
		The top two panels correspond to $|\Delta P_{\mu\mu}|$ and the bottom two panels correspond to $|\Delta P_{\mu e}|$.
		The purple region satisfies this criteria for both methods. In the gray region, our analytic approximations are not valid, since $\alpha \Delta >1$. The white spaces indicate where the analytic approximations are valid but not accurate up to 1\%.
		The small isolated colored patches in the white regions (``islands'') are where the analytic and numerical results match by coincidence. The horizontal dashed lines indicate the baselines $L=1300$ km and $L=7000$ km.
	}
	\label{fig:ContourComparisonpmumu}
\end{figure}

In the previous two subsections, we explored the nature of $P_{\mu\mu}$ and $P_{\mu e}$ at $L=1300$ km, and $L=7000$ km. It was observed that the Expansion method gave very accurate
 estimations at $L=1300$ km, whereas the OMSD approximation gave more accurate results at the baseline of $L=7000$ km. Thus, the values of baselines determine how good a particular approximation would be. In this section, we examine the goodness of the expressions for varying lengths of neutrino long baseline experiments from $L\simeq 30 - 10,000$ km. For these baselines, the neutrinos travel mainly through the curst and mantle of the earth and hence the approximation of constant matter density, which is the average density along that baseline, is expected to work. 
We show our results for energies ranging in 100 MeV -- 30 GeV. Note that we take $\delta_{\text{CP}}=0$, however the accuracy of the expressions would also have a small but finite dependence on the actual values of the $\delta_{\text{CP}}$. 

In figure~\ref{fig:ContourComparisonpmumu} we show the regions in the $(E_\nu,\, L)$ parameter space where OMSD approximation and the Expansion method are able to reproduce the neutrino probabilities to an accuracy of $|\Delta P_{\alpha\beta}| < 1\%$.
Note that since Expansion (with exact dependence on $\gamma_3$) is always expected to formally do better than the Full-Expansion (expanded in $s_{13}$, $\alpha$  and $\gamma_3$) we do not show the results derived with the Full-Expansion method.
It may be observed from figure~\ref{fig:ContourComparisonpmumu} that, at smaller baseline and larger energies, both the analytic approximations give very accurate results. Especially for $P_{\mu e}$ at $L>7000$ km, the OMSD approximation works excellently for energies $E_\nu > 1$ GeV. 

\subsection{The first two oscillation dips in $P_{\mu\mu}$}
\label{sec:oscdip}
We also point out, for a wide range of baselines of neutrino oscillation experiments, one interesting physics observations that could manifests due to the modifications from neutrino decay.
At leading order in the absence of decay, the dips in the oscillation are approximately obtained when $	\Delta =(2n+1) \pi/2$ (where $n=0,1,2,...$). At an oscillation dip, the leading contribution to the survival probability $P_{\mu\mu}$ may be written using eq.~(\ref{eq:pmumuleading}) as
\begin{align}
	P_{\mu\mu}^\text{leading}(\text{dip})=&\;1 - \sin^2 2\theta_{23} -s_{23}^4 \left(1-e^{-4 \gamma_3 \Delta}\right) + 2 s_{23}^2 c_{23}^2 \left(1-e^{-2 \gamma_3 \Delta}\right)\;.
\end{align}
The last two terms are non-zero only when $\gamma_3 \neq 0$. Hence, the deviation of $	P_{\mu\mu}^\text{leading}(\text{dip})$ from $\cos^2 2\theta_{23}$ indicates the presence of neutrino decay. In particular, if $\theta_{23}=45^\circ$, the value of $P_{\mu\mu}$ at the dips would be predicted to be vanishing for the scenario without decay. However, they could be significantly non-zero with decay. 

\begin{figure}[h]\centering
	\includegraphics[width=0.48\textwidth]{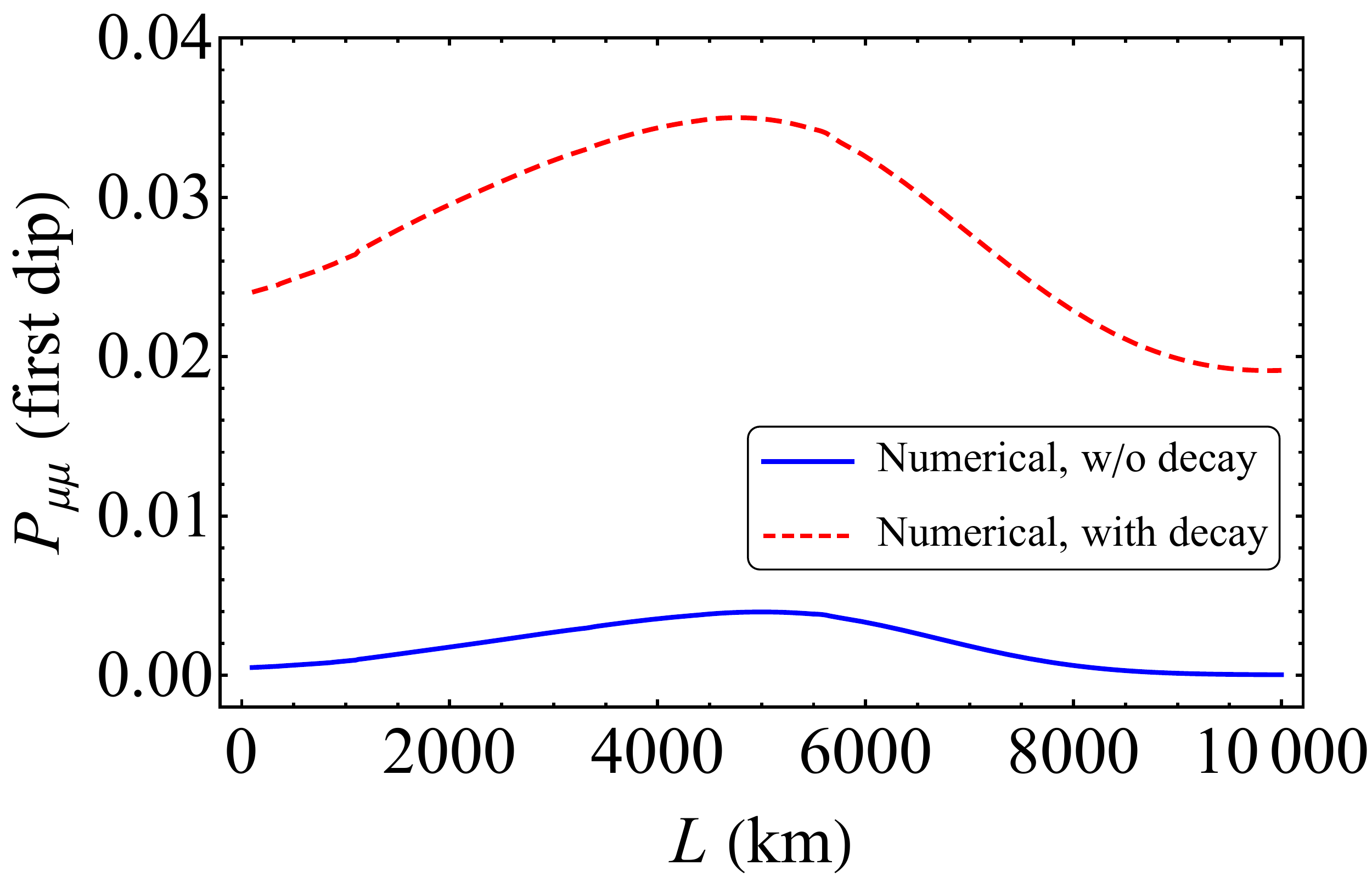}
	\includegraphics[width=0.48\textwidth]{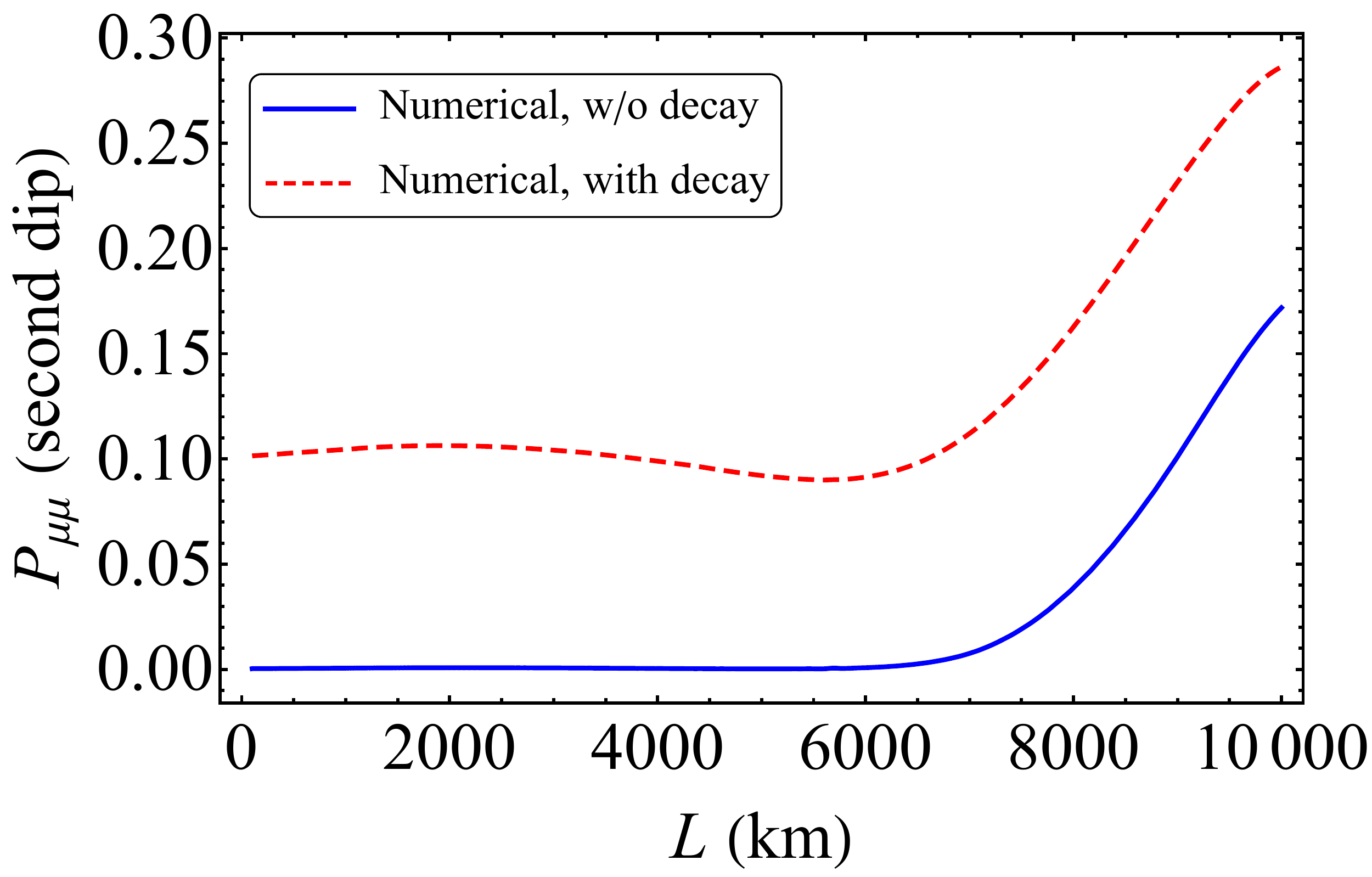}\\
	\caption{
		The survival probabilities $P_{\mu\mu}$ at the first (left) and the second (right) oscillation dips for a range of baselines $L$. We have taken $\theta_{23}=45^\circ$ and $\gamma_3=0.1$. Note that the values of $P_{\mu\mu}$ without decay are non-zero primarily due to Earth matter effects.
	}
	\label{fig:dip}
\end{figure}

At the first oscillation dip, we get
\begin{align}
	P_{\mu\mu} (\text{first dip}) \simeq P_{\mu\mu}^\text{leading} (\Delta \simeq \pi/2) = &\,\frac{1}{4} \left(1-e^{-\pi \gamma_3 }\right)^2 \ge 0 \;,
\end{align}
while at the second oscillation dip,
\begin{equation}
	P_{\mu\mu}(\text{second dip}) \simeq P_{\mu\mu}^\text{leading}  (\Delta \simeq 3\pi/2)  \simeq  \frac{1}{4} \left(1-e^{-3 \pi \gamma_3 }\right)^2 \ge 0\;.
\end{equation}
Both the dips, therefore, are expected to give non-zero values for the survival probability $P_{\mu\mu}$. It is also predicted that the survival probability at the second oscillation dip would be more than that at the first oscillation dip.
For $\gamma_3=0.1$, we obtain from these two equations, $P_{\mu\mu}(\text{first dip}) \simeq 1.8\%$ and $P_{\mu\mu}(\text{second dip}) \simeq 9.3\%$, i.e. an increase of about $\sim 0.02$ at the first oscillation dip, and $\sim 0.1$ at the second oscillation dip.
While this is just a leading order approximation, we show in figure~\ref{fig:dip} the numerical results for a range of baselines, which bear out the predictions regarding the effects of decay extremely well.

This increase in probability may be used as a potent non-trivial signature of neutrino decay. This is especially true for the second oscillation dip, where we get an increase of $\sim 0.1$ in probability in the presence of decay (with $\gamma_3=0.1$). Note that the second oscillation dip is at energies 
\begin{equation}
	E_\nu \simeq 0.69 \left(\frac{L}{1000 \text{ km}}\right)\text{ GeV}\;,
\end{equation}
indicating that its identification would need good energy resolution at lower energies or longer baselines. On the other hand, at an experiment like DUNE the first dip is expected at $\sim 2.7$ GeV, where the flux is sufficiently large and distinguishing the measured survival probability from its prediction without decay may be possible.
\section{Summary and conclusions}
\label{sec:summary}

In this paper, we have presented for the first time, the modifications to the 3-flavor neutrino oscillation probabilities due to possible invisible decay of neutrinos during their propagation through matter. 
We give compact analytic expressions for the probabilities $P_{\mu\mu}$, $P_{ee}$, $P_{e\mu}$ and $P_{\mu e}$, which are relevant for the reactor, long-baseline and atmospheric neutrino experiments.

The inclusion of decay leads to a non-Hermitian effective Hamiltonian, where the Hermitian component is responsible for oscillations, and the anti-Hermitian component gives rise to invisible decay of neutrinos. These two components may not commute in general, leading to a mismatch between the mass eigenstates and the decay eigenstates of neutrinos.
Even if these components commute in vacuum under certain scenarios, they invariably become non-commuting due to matter effects on propagating neutrinos.

Since the constraints on $\nu_1$ and $\nu_2$ decay in vacuum are quite stringent, we specially discuss the scenario where only the $\nu_3$ mass eigenstate in vacuum decays. However, we also treat the most general scenario where the decay matrix $\Gamma$ has all non-zero components. Such a scenario may be relevant in the case of exotic matter effects in the decay of neutrinos.

Our analytic treatment is based on the perturbative expansion in a small book-keeping parameter $\lambda \equiv 0.2$. The small parameters relevant for neutrino oscillations are taken to be $s_{13} \equiv \sin \theta_{13} \sim O(\lambda)$, $\alpha \equiv \Delta m^2_{21}/\Delta m^2_{31} \sim O(\lambda^2)$. The decay parameter $\gamma_3$, based on the current constraints is taken to be $\gamma_3\sim O(\lambda)$. In the most general decay scenario,
the requirement that the decay length-scale should be more than the oscillation length-scale, along with the positive definiteness of the decay matrix, allows $\gamma_1,\; \gamma_2,\; \gamma_{12} \sim O(\lambda^3)$ and $\gamma_{13},\;\gamma_{23} \sim O(\lambda^2)$.

The One Mass Scale Dominance (OMSD) approximation is applicable when the oscillation due to $\Delta m^2_{21}$ can be ignored and only the $\nu_3$ mass eigenstate in vacuum decays. The 2-flavor results from~\cite{Chattopadhyay:2021eba} can then be directly adapted, and the Pauli exponentiation of $2\times 2$ matrices can be implemented. This allows us to calculate the oscillation probabilities with exact dependence on $\theta_{13}^m$, $\gamma_1^m$, $\gamma_3^m$ and $\gamma_{13}^m$, where `$m$' denotes quantities in the presence of matter. 
The probabilities obtained with OMSD approximation have no $\delta_{\text{CP}}$ dependence.
Interestingly, even though we have started with only $\nu_3$ decaying in vacuum, both mass eigenstates in matter, $\nu_1^m$ and $\nu_3^m$, show decaying behavior.
 
The Zassenhaus expansion (inverse Baker-Campbell-Hausdorff), in its resummed version, can be used to incorporate the non-commuting nature of the Hermitian and the anti-Hermitian components of the effective Hamiltonian. This allows the calculation of the probability as a perturbative expansion in the small off-diagonal components of the decay matrix $\Gamma$. The 2-flavor Zassenhaus resummation can be used in conjunction with the OMSD approximation to obtain explicit analytic expressions for the probabilities, exact in $\theta_{13}^m$, $\gamma_1^m$, $\gamma_3^m$, and expanded up to linear order in $\gamma_{13}^m$. When $\theta_{13}^m \approx \theta_{13} \sim O(\lambda)$, these expressions match at $O(\lambda^2)$ with the exact OMSD results obtained using the Pauli exponentiation technique.

We develop the 3-flavor Zassenhaus expansion which expands the applicability of the probabilities beyond the OMSD approximation.
This allows us to calculate the probabilities exact in diagonal elements, and correct up to the first order in the off-diagonal elements, of the decay matrix $\Gamma$.
We find that the effect of $\gamma_3$ on the probability $P_{\mu\mu}$ occurs at $O(\lambda)$, whereas its effect on $P_{ee}$, $P_{e\mu}$ and $P_{\mu e}$ occurs at $O(\lambda^3)$. This indicates that the effects of $\nu_3$ decay would be the most prominent in the $P_{\mu\mu}$ channel. In the most general decay scenario, when all elements of $\Gamma$ are present, the effects of the additional decay terms appear at $O(\lambda^2)$ and higher orders in $\lambda$. Therefore, they are subdominant to the effects of $\gamma_3$, which appear at $O(\lambda)$.
On the other hand, these contributions to $P_{ee}$, $P_{e\mu}$ and $P_{\mu e}$ are at $O(\lambda^3)$, which are at the same order as the leading order contribution due to $\gamma_3$.

In our perturbative calculations, we have expanded in parameters $s_{13}$, $\alpha$, $\gamma_i$, $\gamma_{ij}$ in vacuum, whose values are known to be small. In the presence of matter, one may expect to obtain the corresponding expressions by replacing all parameters with their matter counterparts. However, the values of some of these quantities in matter ($s_{13}^m$, $\alpha_m$, $\gamma_i^m$, $\gamma_{ij}^m$) may not be small. Therefore, for formally correct expansions in orders of $\lambda$, it is desirable to have explicit matter dependence. We achieve this by employing the Cayley-Hamilton theorem which allows the calculations of probabilities in matter in terms of the fundamental quantities $s_{13}$, $\alpha$, $\gamma_i$, and $\gamma_{ij}$ in vacuum, which are definitely small.

In the simpler case where only $\nu_3$ decays in vacuum, we use the Cayley-Hamilton theorem to calculate the neutrino probabilities $P_{\mu\mu}$, $P_{ee}$, $P_{e\mu}$, and $P_{\mu e}$ analytically. These probability expressions are perturbative expansions in $s_{13}$, $\alpha$ and $\gamma_3$, with explicit dependence on the normalized matter potential $A$. They confirm that even in the presence of matter, the effect of $\nu_3$ decay manifests at $O(\lambda)$ in $P_{\mu\mu}$, whereas for  $P_{ee}$, $P_{e\mu}$, and $P_{\mu e}$, the effects of decay occurs at $O(\lambda^3)$. Moreover, the decay contribution to $P_{\mu\mu}$ is independent of matter effects up to $O(\lambda^2)$. For $P_{ee}$, $P_{e\mu}$, and $P_{\mu e}$, we observe that the decay contribution has matter dependence, with $P_{e\mu}$ and $P_{\mu e}$ having the $\sin^2[ (A-1)\Delta] / (A-1)^2$ dependence similar to the terms without decay.
We extend our calculations further to include the exact dependence on $\gamma_3$ (with expansion in $s_{13}$ and $\alpha$) to point out the phenomenologically rich functional dependence of the probabilities on $\gamma_3$, in addition to the naively expected exponential decay terms.

In the general case where all the elements of the decay matrix $\Gamma$ may be non-zero, we calculate the additional contributions due to $\gamma_1$, $\gamma_2$, $\gamma_{12}$, $\gamma_{13}$ and $\gamma_{23}$. We observe that even in matter, the effects of these additional decay elements on $P_{\mu\mu}$ appear at $O(\lambda^2)$ or less, compared to the effects of $\gamma_3$, which appear at $O(\lambda)$. Thus, these off-diagonal terms will not significantly affect the signatures of decay in $P_{\mu\mu}$. As observed earlier, for $P_{ee}$, $P_{e\mu}$, and $P_{\mu e}$, the effects of these elements remain as important as the $\gamma_3$ contribution, all appearing at $O(\lambda^3)$ even in matter.

Our analytic observations have helped us get insights into the impact of possible neutrino decay on the probabilities and estimate the extent of this impact qualitatively in terms of an expansion in $\lambda$. Finally, we compare our analytic expressions with the exact numerical results, calculated for long-baseline experiments. For illustration, we take $L=1300$ km, which is close to the baseline of the Deep Underground Neutrino Experiment (DUNE) and $L=7000$ km, which corresponds to the ``magic baseline''. Our comparison of $P_{\mu\mu}$ and $P_{\mu e}$ shows that all our analytic approximations reproduce the peak and dip positions in the probability quite accurately. As far as the values of the probability are concerned, the ``Expansion'' (obtained using the Cayley-Hamilton theorem) is a very accurate analytic approximation at $L=1300$ km, with $|\Delta P_{\alpha \beta}| <0.01$ at energies more than a few GeV. The OMSD approximation gives results with such accuracy at $L=7000$ km.
The expected leading order effects of the decay, at $O(\lambda)$ on $P_{\mu\mu}$, are confirmed with the numerical calculations.
We show the regions in the $(E_\nu, L)$ parameter space where our two major analytic approximations, viz. OMSD approximation and Expansion, give results correct to $|\Delta P_{\alpha\beta}|<0.01$. These indicate that our approximations work very well for lower values of $\Delta=\Delta m_{31}^2 L/(4E_\nu)$.

Analytic approximations often reveal hidden features of interest which may not be immediately apparent while looking at numerical results. We observe such features in the heights of the dips in the muon neutrino survival probability $P_{\mu\mu}$. Our analytic approximations, at their leading order, suggest the non-intuitive result that the probabilities at these dips would have \textit{higher} values in the presence of decay, compared to those in the absence of decay. This conclusion is born out by the exact numerical calculations and stays valid even in the presence of matter at long baselines. The increase in this probability may be as much as $\sim0.02$ at the first oscillation dip, and $\sim 0.1$ at the second oscillation dip, for $\gamma_3=0.1$. This feature may be useful as a unique sensitive signature to confirm the presence of possible neutrino decay, or put strong constraints on it.

\section*{Acknowledgments}
D.S.C. and A.D. acknowledge support from the Department of Atomic Energy (DAE), Government of India, under Project Identification No. RTI4002.
K.C. acknowledges the support from Shanghai Pujiang Program (20PJ1407800) and National Natural Science Foundation of China (No. 12090064). S.G. acknowledges the J.C Bose Fellowship (JCB/2020/000011) of Science and Engineering Research Board of Department of Science and Technology, Government of India.

\providecommand{\href}[2]{#2}\begingroup\raggedright\endgroup

\end{document}